\documentclass[preprint,journal]{vgtc}       

\ifpdf
  \pdfoutput=1\relax                   		
  \usepackage{graphicx}                		
  \DeclareGraphicsExtensions{.pdf,.png,.jpg,.jpeg} 
\else
  \usepackage{graphicx}                	  
  \DeclareGraphicsExtensions{.eps}     
\fi%

\graphicspath{{figures/}{pictures/}{images/}{./}}

\usepackage{microtype}                 		
\PassOptionsToPackage{warn}{textcomp}       
\usepackage{textcomp}                  		
\usepackage{mathptmx}                  		
\usepackage{times}                     		
\usepackage{cite}                      		
\usepackage{tabu}                      		
\usepackage{booktabs}                  		

\usepackage{amsmath}
\usepackage{amssymb}
\usepackage{bbm}
\usepackage{xcolor}							

\newif\ifproofread
\newcommand{\changemarker}[1]{%
\ifproofread
\textcolor{red}{#1}%
\else
#1%
\fi
}

\usepackage{tikz}
\usetikzlibrary{fit}
\usetikzlibrary{shapes,arrows}
\usetikzlibrary{fadings}

\newcommand \mbl[1]{%
    \tikz[overlay,remember picture]
        \node (marker-#1-a) at (0,1.2ex) {};%
}
\newcommand \mbr[2]{%
    \tikz[overlay,remember picture]
        \node (marker-#1-b) at (0,0.3ex) {};%
    \tikz[color = #2, overlay,remember picture,thick,inner sep=3pt]
        \node[draw,rounded rectangle,fit=(marker-#1-a.center) (marker-#1-b.center)] {};%
}

\onlineid{0}

\vgtccategory{Research}
\vgtcpapertype{please specify}

\title{Audio-Visual-Olfactory Resource Allocation for Tri-modal Virtual Environments}

\author{E. Doukakis, K. Debattista,  T. Bashford-Rogers, A. Dhokia, A. Asadipour, A. Chalmers, and C. Harvey}
\authorfooter{
\item Correspondence to E.Doukakis, stratosver2@gmail.com or K.Debattista, k.debattista@warwick.ac.uk
\item E. Doukakis, K. Debattista, A. Dhokia and A. Chalmers are with University of Warwick
\item A. Asadipour is with Computer Science Research Centre at RCA
\item T. Bashford-Rogers is with University of the West of England
\item C.Harvey is with Birmingham City University
}

\shortauthortitle{Firstauthor \MakeLowercase{\textit{et al.}}: Paper Title}

\proofreadfalse		

\abstract{Virtual Environments (VEs) provide the opportunity to simulate a wide range of applications, from training to entertainment, in a safe and controlled manner. For applications which require realistic representations of real world environments, the VEs need to provide multiple, physically accurate sensory stimuli. However, simulating all the senses that comprise the \changemarker{human sensory system (HSS)} is a task that requires significant computational resources. Since it is intractable to deliver all senses at the highest quality, we propose a resource distribution scheme in order to achieve an optimal perceptual experience within the given computational budgets. This paper investigates resource balancing for multi-modal scenarios composed of aural, visual and olfactory stimuli. Three experimental studies were conducted. The first experiment identified perceptual boundaries for olfactory computation. In the second experiment, participants ($N = 25$) were asked, across a fixed number of budgets ($M = 5$), to identify what they perceived to be the best visual, acoustic and olfactory stimulus quality for a given computational budget. Results demonstrate that participants tend to prioritize visual quality compared to other sensory stimuli. However, as the budget size is increased, users prefer a balanced distribution of resources with an increased preference for having smell impulses in the VE. Based on the collected data, a quality prediction model is proposed and its accuracy is validated against previously unused budgets and an untested scenario in a third and final experiment.%
}


\begin{document}

\firstsection{Introduction}
\maketitle

The existence of multiple stimuli in a VE is required for increasing immersion in many current and future applications. The inclusion of realistic olfactory delivery along with audio-visual stimuli is of key importance if VEs are to be used as genuine representations of real life scenarios \cite{NakYos06}. The introduction of smell impulses increases the sense of presence in the virtual world and enhances the level of realism \cite{Chen06}. The significance of a tri-modal combination of smell, vision, and hearing might not be initially obvious but it affects how people think, feel and behave \cite{SSP04}.

Accurate computation and delivery of multiple stimuli in high fidelity requires significant computing capability. Previous research has shown that humans cannot fully attend to all the incoming sensory stimuli in the real environment. Such cross-modal interaction phenomena and known limitations of the HSS have been utilized to reduce computational requirements \cite{Hulusic2012} without the users being able to perceive any quality degradations.
Furthermore, it is unclear how best to allocate computational resources in a VE. For example, if an improvement in computational performance of $50\%$ becomes available, how would it be best to make use of that extra computational power. Is it best spent on improving visuals? Improving the audio? A smaller improvement to both? Adding smell? Resource allocation schemes have been recently proposed to describe distribution of available resources based on human subjective preferences \cite{Doukakis2018}. However, previous work is limited to two modalities, in particular, audio and vision.

The work presented in this paper captures how human allocation preferences are adjusted given a budget of computational resources in a tri-modal VE set up. In order to establish how this computational budget can be distributed between the three senses, three experiments have been conducted. Experiment I was focused on understanding the perceptual limitations of olfactory computation, such that the computational requirements for olfactory delivery can be established. With this knowledge, Experiment II was conducted whereby participants allocated computational resources across visual, auditory and olfactory stimuli in order to identify a perceptually optimal load balancing of these resources from a given fixed budget. This was conducted at five different computational budgets for a number of scenarios. Based on the collected subjective data, a resource distribution model is proposed and evaluated on untested budget sizes and scenarios in Experiment III. Therefore, multi-sensory rendering pipelines can exploit such a model to direct resource allocation decisions in VEs.

The main contributions of this work are as follows:
\begin{itemize}
\item A psychophysics framework for estimating odour \changemarker{just noticeable difference (JND)} thresholds for a range of smell concentration magnitudes.
\item An experimental methodology for allocating resources in tri-modal VEs.
\item Evidence that participants generally prefer to allocate resources for visual stimuli. However, as the budget increases the percentage devoted for aural and olfactory stimuli in the virtual scenario is increased significantly.
\item A validated model capable of predicting resource allocation in systems where visual-aural and olfactory cues are intended for delivery.
\end{itemize}

\label{sec:introduction}

\section{Background and Related Work}\label{sec:background}
Simulation and delivery of multiple senses at the same time is considered crucial for ensuring a realistic experience and increase a user's overall level of immersion~\cite{VRDurlach}. Applications of multi-sensory VEs range across different sectors of academia and industry. In reality, perceiving one sensory stimulus is quite rare in the physical world and many studies that investigate multiple senses, have found that the perceptual impact of one sense to the other can be quite significant~\cite{Bertelson04,Driver1998}.

\subsection{Audio-Visual Interactions}
Increases in reported presence have been found in audio-visual VEs across a range of acoustic conditions~\cite{HB96, LVK02, BA*17}. Indeed, overall quality perception in audio-visual environments is mediated by the quality in each sense \cite{storms98}. This showed that better audio increases the perception of quality in the visual domain whilst the opposite was also observed: better visuals decreased the perception of quality in the acoustic domain. This effect was shown to be practical by Moeck \emph{et al.} by using hierarchical clustering of sound sources given congruent visual signals \cite{Moeck07}. \changemarker{Perceptual interactions as triggered by sound cues are explored by Rocchesso~\cite{Rocchesso11}. Using sound effects, a series of human-human and human-object interactions are explained and validated. Preserving the level of presence in a VE has recently been examined by Grani~\emph{et al.}~\cite{Grani14}. In their work, audio-visual attractors are used in an experimental study to quantify how users' attention is directed in a cave automatic virtual environment, avoiding gaps in presence.} Sound has also been shown to influence perception in the spatial domain, such that congruent sounds can direct attention. This effect has been used in selective rendering models~\cite{HD*16}.

\subsection{Olfactory-Visual Interactions}
Olfactory-visual stimulation has been shown to increase presence in both generalized virtual environments and in targeted virtual environments when compared to a visual only condition \cite{DW*99, MNBJ16}. Supplementary to this, Munyan \emph{et al.} \cite{MNBJ16}, showed that when the olfaction condition was removed, this resulted in a disproportionate decrease in presence. Attentional changes have been observed when humans are presented with multi-sensory stimuli when compared to a visual only condition. Seo \emph{et al.}~\cite{SRM*10} observed congruent objects impacting viewing time and deviating eye fixation. Seignuric \emph{et al.}~\cite{SDJ*10} investigated the influence of \emph{a-priori} connections in between a scent and congruent visual stimuli on eye saccades and fixations, showing that congruent objects were explored faster in the presence of the odour. Chen \emph{et al.} \cite{Chen13} performed a study to corroborate this effect and concluded that a multi-modal saliency map weighing both visual and olfactory inputs was required. Harvey \emph{et al.} \cite{HB*18} showed that conventional image saliency maps can no longer be relied upon in the same way in olfactory-visual environments and demonstrated a validated model based on empirical findings.

\subsection{Multisensory Integration}
Burr and Allais have proposed a linear model for bimodal fusion in the audio-visual domain \cite{Burr2006}. This suggests that weights control the bimodal information from the two senses: $\hat{S} = w_A \hat{S_A} + w_V \hat{S_V}$, where $w_A$ and $w_V$ scale the estimates for audio and vision respectively, $\hat{S_A}$ and $\hat{S_V}$. Multisensory VEs are computationally demanding when considering the simulation of numerous senses \cite{GTLG14}. It is however possible to balance computation to account for the weight that the human sensory system places on each sense. However, multisensory VEs have inherent perceptual affects that have to be understood \cite{AMMG16}, before these weights can be derived. In the study proposed by Doukakis \emph{et al.} \cite{Doukakis2018}, the authors presented a method for resource allocation in bi-modal VEs, namely vision and hearing, based on human subjective preferences. In that experimental study, participants allocated a given budget of resources to improve the quality of the audio-visual stimuli. Based on the results, an estimation model is proposed and validated. \changemarker{Similar methods have been used by Slater \emph{et al.}~\cite{Slater10} to investigate the level of presence in VEs by conducting experiments where experienced participants at immersive system methodologies vary four possible graphical factors.
In a subsequent experimental study by Skarbez \emph{et al.}~\cite{Skarbez17}, participants could adjust a series of coherence factors to increase the level of the plausibility illusion, to match the perceptual experience they had in a highest coherence scenario. Results showed that participants prioritize improvements to the virtual body.}

\changemarker{Ernst and Banks~\cite{Ernst02} investigated which of the senses of vision and haptics is more dominant using a maximum-likelihood estimation on the combined input of both sensory cues. Using the variances of each sense in height estimation, a maximum-likelihood integrator model is given and compared to human collected data in visual-haptic tasks. Azevedo \emph{et al.}~\cite{Azevedo14} considered how the senses of vision, hearing, olfactory and haptics are classified for measuring presence, focusing on outdoor VEs. Their results showed that the combined effect of haptics and hearing was considered more important than the typical VE stimuli of vision and hearing, dependent on scene and plausibility illusions.}

\changemarker{In summary, there exists a large body of work that considers the permutations of senses and their respective influences on human perception and presence. Studies that consider multisensory integration are shown to be of benefit in VEs when resources are adapted based upon empirical findings.
In the bi-modal case, this resource allocation has been quantified but beyond remains unexplored. }


\section{Motivation and Overview}
In this work, we are interested in identifying how to best allocate computational resources across audio, vision and olfaction. While audio and vision are reasonably well understood and have been used in a significant number of cross-modal experiments \cite{Hulusic2012}, this is not the case with olfaction. Following the audio-visual approach of Doukakis et al. \cite{Doukakis2018}, our tri-modal model is built around permitting users to adjust the required computation for all the senses to fit within a given computational budget. The selection and adjustment of the aural and visual stimuli is based on the approach adopted by Doukakis et al. \cite{Doukakis2018}. However, since the application of olfaction in virtual environments is less understood, our initial experiment (Experiment I, Section \ref{sec:preexp}) seeks to identify a useful perceptual parameterization for olfactory stimuli. Experiment II (Section \ref{sec:expFramework}) then collects data for load balancing across the tri-modal stimuli at five budgets across three scenarios. Section \ref{sec:models} uses these results to develop a model for tri-modal resource allocation. Finally, Experiment III (Section \ref{sec:Validation}) validates the model. 

\section{Experiment I: Identifying olfactory parameters}\label{sec:preexp}
This section describes the olfactory simulation and its parametrisation. Olfaction will be parameterized by the mesh size (as we are using a finite element solver) as it effects the convergence rate of our olfactory simulation using \changemarker{computational fluid dynamics (CFD)}. An experimental framework is then described for estimating JNDs in a range of concentration magnitudes. The JND threshold estimation allows us to assess whether the recorded concentration curves for different mesh discretization levels are perceptually equivalent. As shall be shown, simulations across a wide range of mesh parameterizations (from 1K to 1M) are not perceptually noticeable by human observers.

\subsection{Simulation}

\begin{figure*}[!h]
  \centering
  \includegraphics[scale=0.2]{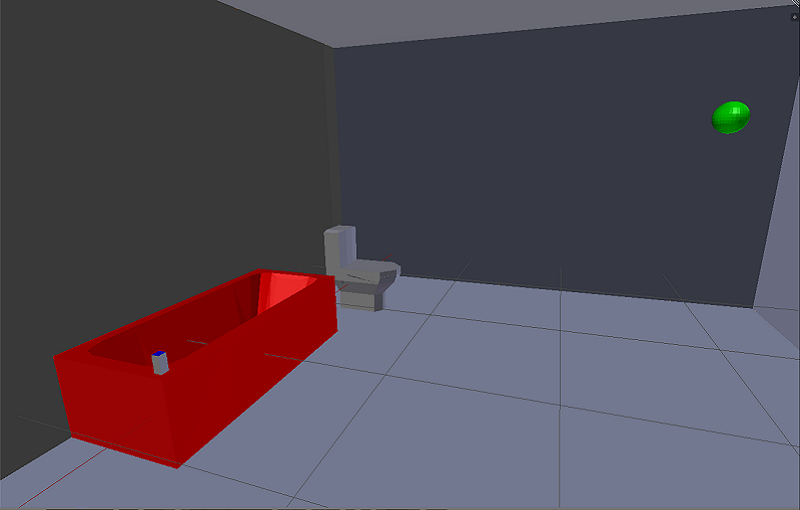}
  \hspace{0.5mm}
  \includegraphics[scale=0.2]{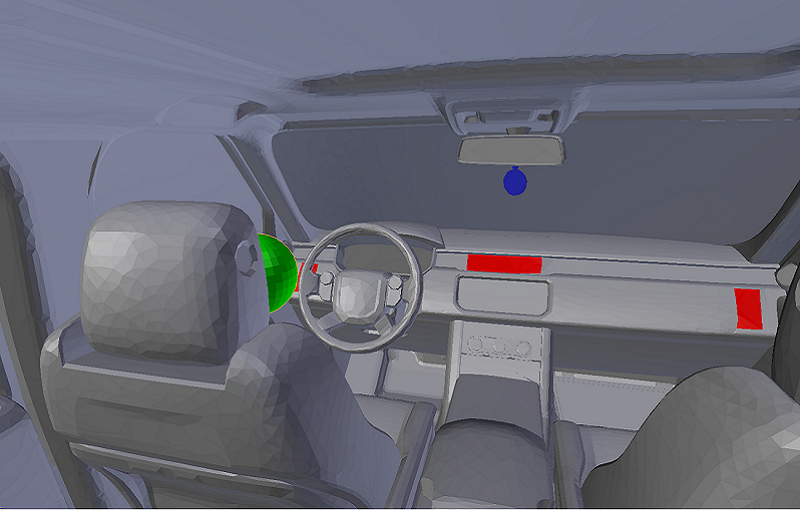}
  \hspace{0.5mm}
  \includegraphics[scale=0.2]{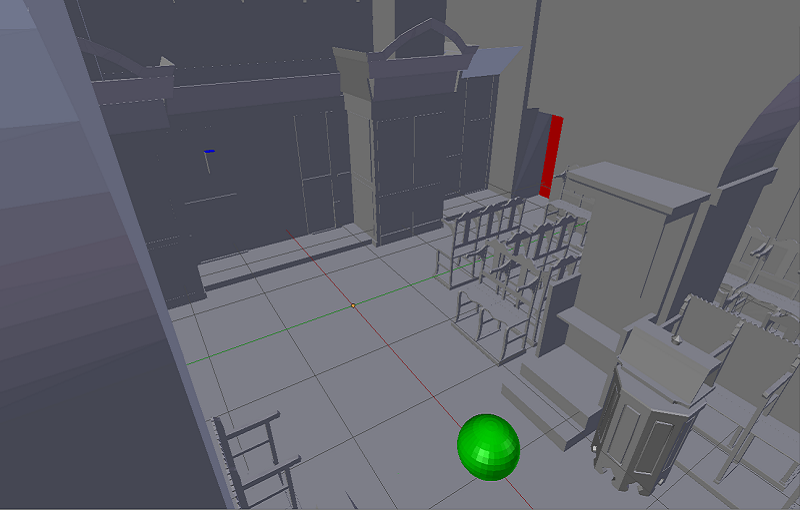}
  \hspace{0.5mm}
  \includegraphics[scale=0.2]{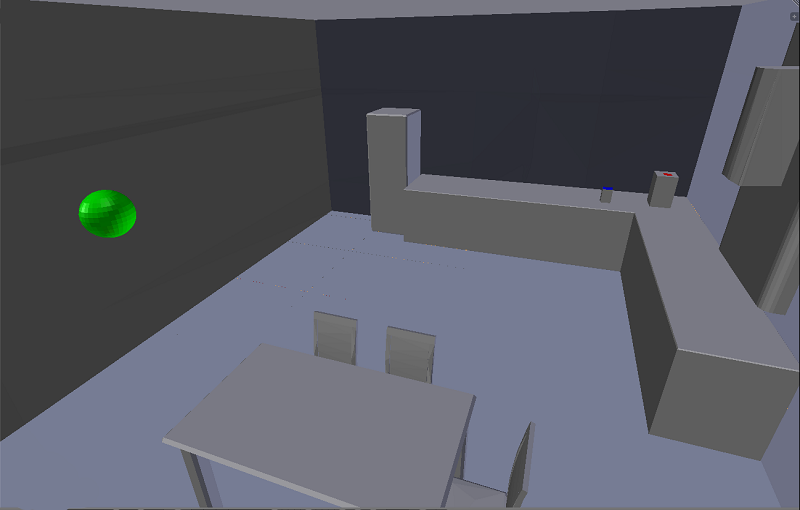}

  \caption{The boundary conditions of the VEs used at this experimental study, From left to right: Bathroom interior, Car, Kitti, Kitchen. The blue painted patches represent the odour inlets while the red patches are used for developing convention effects. The green coloured spheres represent the virtual probes for reading concentration values.}
  \label{fig:VE-Boundaries}
\end{figure*}

This section presents the framework used for simulating smell transport in four test VEs, \changemarker{as shown in Figure \ref{fig:VE-Boundaries}}, using CFD. Odour concentration is captured at virtual probes during the simulation stage across a range of different quality meshes. In this work, the considered mixture is composed by the smell of citral in air and the objective is to estimate the concentration of citral during the transport of the mixture.

All the CFD simulations were implemented using the \textit{Eulerian} approach, i.e. compute the variables of interest over time at the centres of the \changemarker{control volumes (CVs)} that compose the domain's computational mesh. This procedure requires two steps. Firstly, it includes the discretisation and solution of the Navier-Stokes equations that govern the transport of the air-odour mixture and compute its momentum ($\overrightarrow{U}$), pressure ($p$) and density ($\rho$). Secondly, coupling the governing equations with a transport differential equation for estimating the concentration of the odour, denoted as $C$, at every node of the mesh. This equation is of the general form:
\begin{equation}
\underbrace{\frac{\partial \rho C}{\partial t}}_{\text{temporal term}} + \underbrace{\nabla \cdot (\rho \overrightarrow{U} C)}_{\text{convection term}} - \underbrace{\nabla \cdot \overrightarrow{J}}_{\text{diffusion term}} = \underbrace{S_{C}(C)}_{\text{source term}},
\label{transportEquation}
\end{equation}
where the quantity $\overrightarrow{J} = \rho D_{ca} \nabla C)$ is the mass diffusion flux of the odour and $D_{ca} = 8.23 \times 10^{-5}$ m$^2$/s is the diffusion coefficient of citral into the air~\cite{Mar72}. The term $S_{C}(C)$ describes the effect of body forces and is used to model the effect of the Earth's gravitational field. The temporal term and the terms that model the physical processes of convection and diffusion are underlined in Equation~\ref{transportEquation}. As is the case with every odorous gas, the flow transport of the citral-air mixture is highly turbulent. Turbulence effects were implemented using a realisable $k-\epsilon$ model~\cite{Tsan95}.

The domain's mesh granularity affects the solution's convergence rate while it can increase numerical stability during the simulation \cite{Sande14}. Coarse meshes yield concentration solutions with over- and undershoots because of the non-smooth transition of the odour-air mixture across the CVs. We considered four different mesh versions of successively higher number of CVs for every one of the three scenarios. These are meshes with $1$K, $10$K, $100$K and $1$ million CVs. The coarsest mesh was refined near the surfaces and the boundaries of the VE for better accuracy while every successive refinement was uniform across the domain so as to approximately preserve the initial distribution of CVs in the boundaries. Mesh sizes with an order of magnitude change in the number of CVs were selected in order to study how odour concentration at probe locations changes between large spatial discretisation steps starting from very coarse up to excessively high refinement levels.

\subsubsection{Application to VEs}
\label{sec:VEs}
Simulation of smell propagation was considered in four different VEs. These are depicted in Figure~\ref{fig:VE-Boundaries}. All the physical quantities used as boundary conditions were obtained by measuring concentration and flow rate with a photoionization detector and a flow meter respectively in the real places that were used for creating the VEs. The final values resulted through averaging of the results collected over $10$ repeated measurements. These scenarios were chosen because the convection process occurs in a different way in each scenario, and therefore the smell will be dispersed differently. In the Bathroom, temperature differences between the hot bath ($45$ $^{\text{o}}$C) and the environment ($15$ $^{\text{o}}$C) cause the smell-air mixture to circulate in the room. In the car scenario, convection is simply created due to the air flow coming through the vents at a constant temperature of $25$ $^{\text{o}}$C. In the Kitti scenario, convection occurs due to the air coming through the three doors. In the Kitchen scenario, convection effects are introduced through the temperature gradient between the hot kettle ($50$ $^{\circ}$C) and the cold environment ($15$ $^{\circ}$C). Pressure was assumed to be the same and equal to atmospheric pressure ($101.325$ Pa) for all three VEs. Figure~\ref{fig:VE-Boundaries} depicts the smell inlet boundaries and the patches used for developing convection effects in the VEs. The same figure also shows the locations of the virtual probes in the VEs.


\subsubsection{Smell concentration results}
Smell concentration values were recorded at the virtual probe using a sampling frequency of $4$Hz. This sampling rate was chosen based on the rate many photoionisation devices record odour concentration values in real environments \cite{HW06-PID}.

\begin{figure}[!ht]
\centering
\includegraphics[scale=0.27]{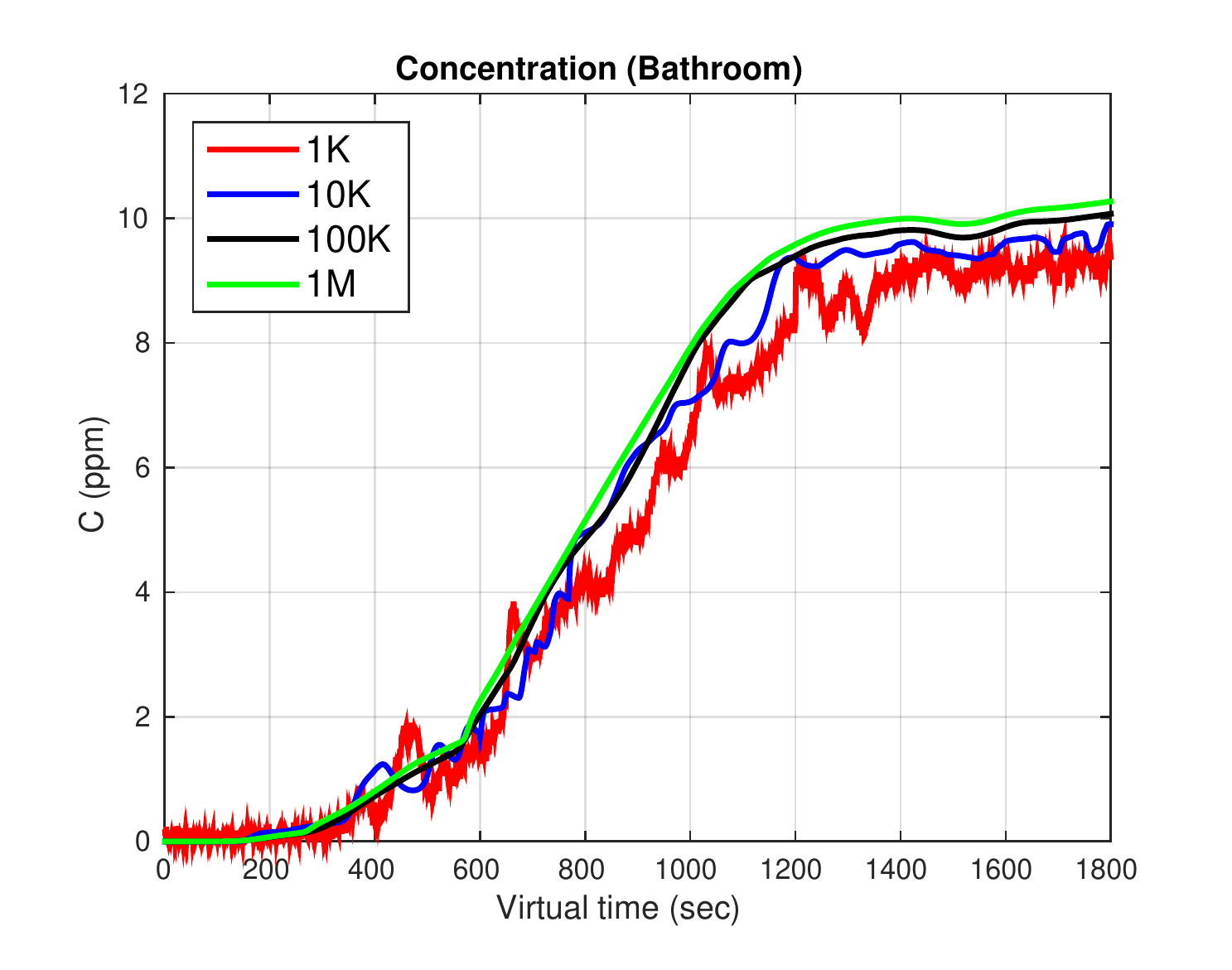}
\includegraphics[scale=0.27]{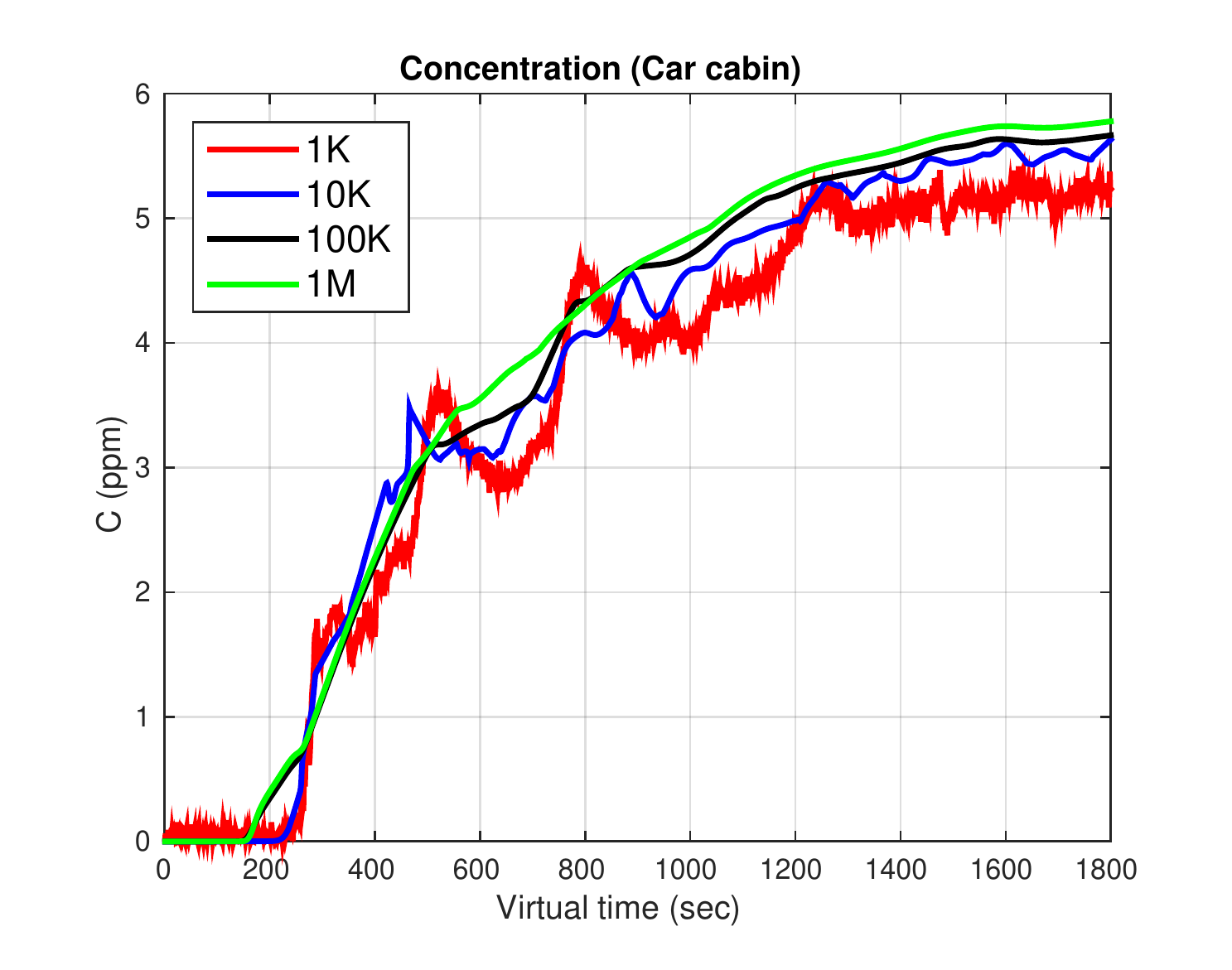}\\
\vspace{0.1mm}
\includegraphics[scale=0.26]{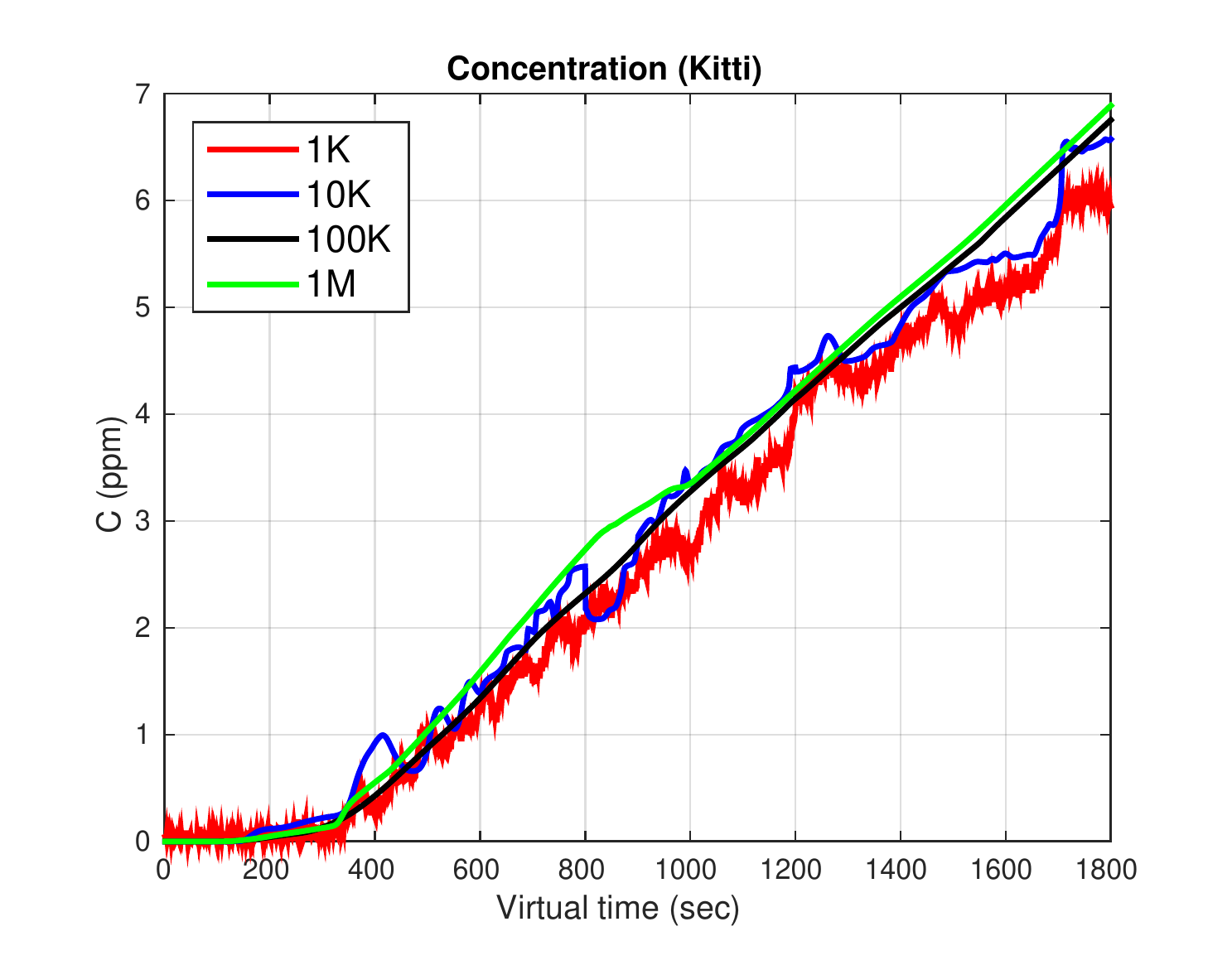}
\includegraphics[scale=0.27]{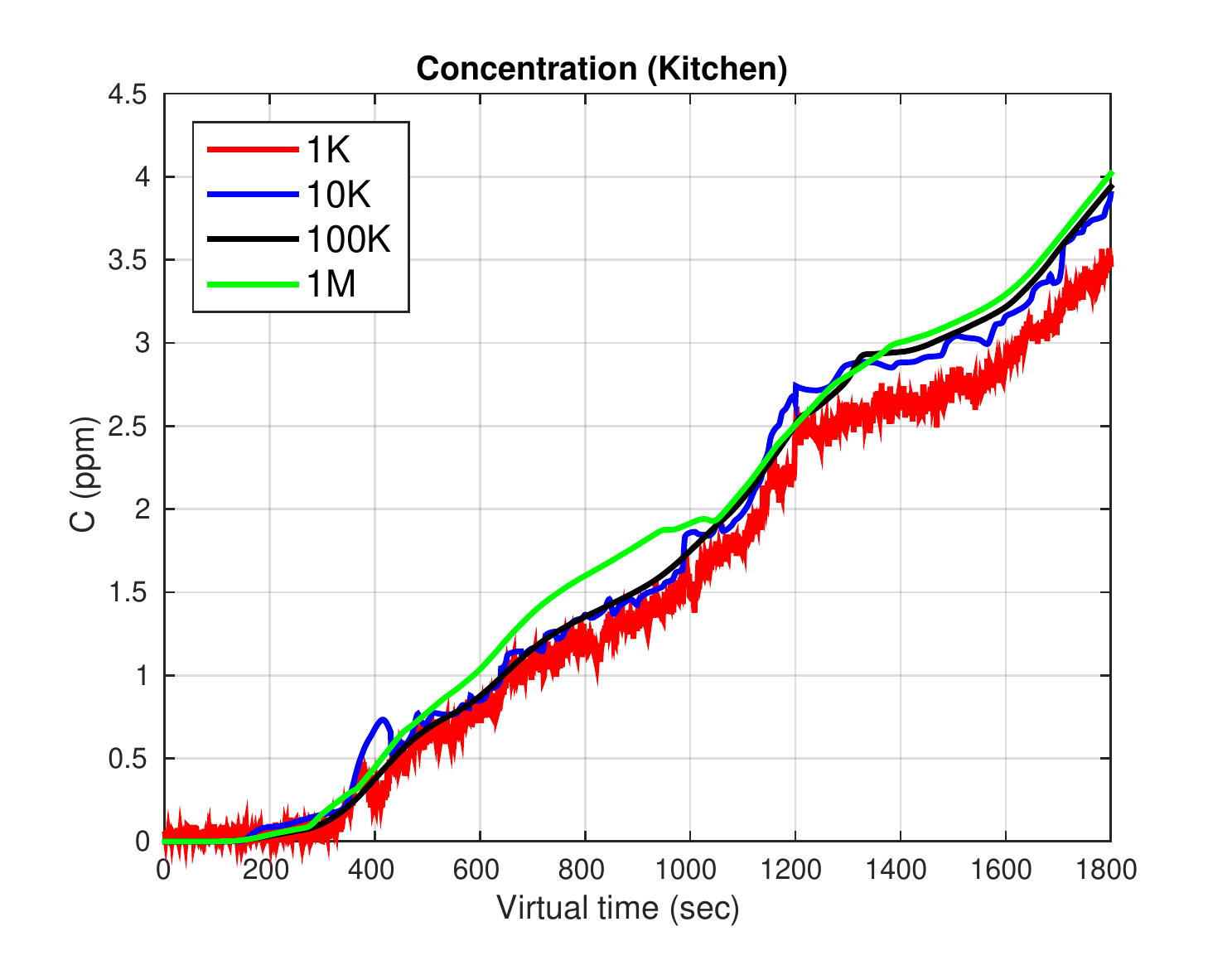}\\
\caption{Concentration results measured for different mesh discretisation levels at the virtual probe locations. Top-left: Bathroom, Top-right: Car, Bottom-Left: Kitti, Bottom-Right:Kitchen.}
\label{fig:concentration}
\end{figure}

For all the VEs, smell propagation was simulated for $1,800$ seconds of virtual time to see how concentration evolves at the sensor location for a sufficient amount of time. Figure~\ref{fig:concentration} shows how concentration changes over time for the four different spatial discretisation levels. For coarse meshes the transition of the air-citral mixture near the sensor position is more abrupt therefore fluctuations are captured in the concentration function. These fluctuations are progressively eliminated as the mesh is refined. As can be seen from the graphs in Figure~\ref{fig:concentration}, the concentration curves are not very distant from each other. Specifically, the maximum difference between the coarsest (1K) and the finest meshes (1M) was \changemarker{$1.98$ ppm} for the Bathroom scenario, $1.623$ ppm for the Car scenario and \changemarker{$1.345$ ppm} for the Kitti scenario. Furthermore, the curves stabilise as the VE becomes saturated from the emitted odour. After this equilibrium point, concentration does not change significantly. The equilibrium point has not been fully achieved for the Kitti scenario because of the large volume of the room ($560.64$ m$^3$) and the location of the sensor.

\subsection{Method}
In this study the experimental methodology used for estimating the JND threshold was the two interval forced choice (2IFC) method~\cite{KP09}. At every trial two olfactory stimuli were presented with an interstimulus time interval (ITI) between them. The first is a constant concentration stimulus, known as a pedestal, while the second is a varying stimulus randomly selected from a predefined set of concentration magnitudes. Ten trials were implemented for each comparison pair (pedestal, varying) for the different values of the varying stimulus. The presentation order of the two smell impulses (pedestal-varying or varying-pedestal) was also randomized across the experimental trials to avoid any potential ordering bias.

The JND threshold is the magnitude that needs to be added to the pedestal so as the resulting olfactory stimulation has a magnitude that is perceptually more intense than the pedestal in $75 \%$ of the total trials~\cite{KP09}. During each trial, the two stimuli were presented with an ITI interval of $4$ seconds while the stimulus exposure duration (SED) was $0.5$ seconds for each of the two delivered bursts. A short SED was selected to avoid user adaptation to the released smell while the ITI was sufficiently long to permit the participants to rest before inhaling the second stimulus.

In this work, a computer-controlled olfactory display was used to deliver the smell impulses at the two SEDs~\cite{DDAHBDWC16}. This device provides two independent channels that can release smell impulses synchronously while it contains two digital mass flow controllers (DMFCs) for adjusting the flow rate of each channel as a proportion of the maximum flow rate provided which is $1000$ ml/min. For our olfactory display, the suggested operating range for guaranteed concentration accuracy of the delivered stimulus is $I = [C_{0}, C_{\text{max}}]= [1.2, 11.2]$, parts per million (ppm) for a SED time of $0.5$ sec. All the participants were able to detect the lemon like odour when they sniffed $C_{0}$, thus, the whole range of stimuli contained in $I$ were higher than the absolute detection threshold.

As there was no previous knowledge for the existence of any JND threshold in the range $I$, starting with $C_{0}$ as the pedestal, pairwise comparisons of $C_{0}$ with a number of varying stimuli that uniformly discretize $I$ were planned until the average performance among all the participants would reach a proportion of $75 \%$.  For a step $s$ the interval $I$ was subdivided to three mutually exclusive subsets:
\begin{equation}
\begin{split}
I_{1} & = [1.2 + s, 1.2+2s, \cdots, 1.2+8s]\\
I_{2} &= [4.8 + s, 4.8+2s, \cdots, 4.8+8s]\\
I_{3} &= [8 + s, 8+2s, \cdots, 8+8s].
\end{split}
\label{eq:intervals}
\end{equation}
A step size $s$ equal to $0.4$ was selected so as to keep a reasonable number of varying stimuli for each of the three subintervals and be able to span the whole range in the extreme case that no perceptually different magnitude existed between $C_{0}$ and any of the stimuli in $I_{1}, I_{2}, I_{3}$. Starting with comparisons of the pedestal $C_{0}$ with the stimuli in interval $I_{1}$, if there was no stimulus that triggers a perceptual difference, denoted as $C_{1}$ the experimentation would be carrying on in a subsequent experimental session for the same pedestal and the stimuli with magnitudes in the interval $I_{2}$ until the first stimulus that triggers a perceptual difference. If there was no such stimulus, the pedestal would be compared with stimuli in $I_{3}$. In case $C_{1}$ could be found, there is no need to continue experimenting for the rest of the stimuli in any of the $I_{2}$ and/or $I_{3}$ as it was necessary to find perceptual differences starting from $C_{1}$ as the new pedestal.

The first time that a JND threshold was found a new experimental session was initiated by adjusting the step for spanning the rest of the interval $I$ using Weber's law. This law has been applied in the olfactory domain~\cite{AG98} and states that the ratio of the JND to the pedestal stimulus is constant. Specifically:
\begin{equation}
\frac{\text{JND}}{C_{p}} = \frac{C_{f} - C_{p}}{C_{p}} = k,
\label{eq:Weber}
\end{equation}
where $C_{p}$ is the pedestal and $C_{f}$ the final intensity stimulus that triggers a perceptual difference. The constant $k$ can be obtained from the first JND, denoted as JND$_{1}$, and the first pedestal $C_{0} = 1.20$ ppm. Using the constant $k$, we can obtain an estimation of the final stimulus that elicits a perceptual difference starting from $C_{1}$ as the new pedestal and applying again Weber's law. In case the estimation is not in the interval $I$, no further experimentation is required as it cannot be delivered by the olfactory display. If not, a subinterval with a new discretization step is derived and a new experimental session is initiated. The new subinterval includes Weber's law estimation as the potential stimulus that elicits perceptual difference. The new step is computed so as the new interval does not exceed $10$ varying concentration magnitudes.

This adaptive method allowed us to speed up the process of finding a new stimulus that triggers a perceptual difference without comparing intermediate pairs that would be perceptually similar using the initial small step of $0.4$. The derivation of a new step enabled avoiding unnecessary experimental trials. Using this methodology, two JND thresholds were found in the operating range $I$. The first one was included in $I_{1}$ while the second one was found after one step adjustment using an interval of $10$ varying stimuli.

\subsubsection{Materials}
\begin{figure}[htb]
\centering
\includegraphics[scale=0.3]{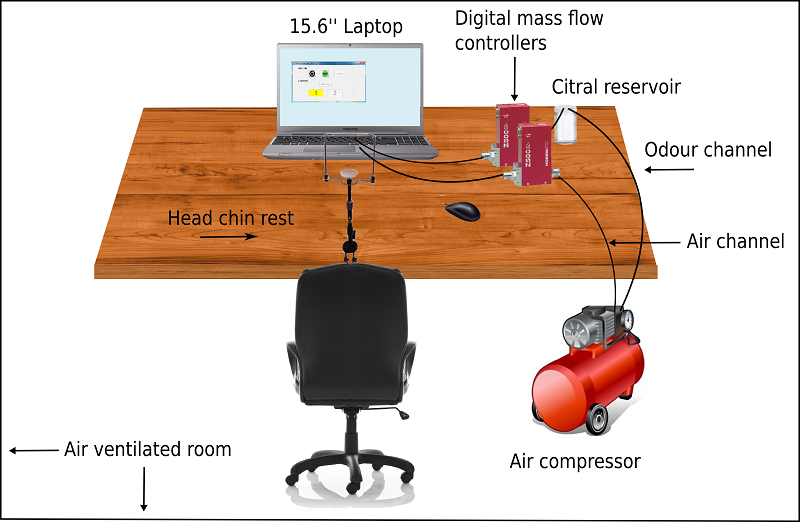}
\caption{Experimental set up for Experiment I.}
\label{fig:JNDSetup}
\end{figure}

For this study a $15.6^{\prime \prime}$ laptop computer was used for controlling the olfactory display and the \changemarker{graphical user interface (GUI). The apparatus is shown schematically in Figure~\ref{fig:JNDSetup}.} Two tubes arrived near the nose from the olfactometer, one that delivered smell bursts at every experimental trial and one air control channel that was evacuating the old smell stimuli by releasing humidified air at an exact flow rate of $1000$ mL/min during the ITI.

The distance of the nose away from the smell delivery point was kept constant at $7$ cm using a chin rest located in front of the laptop. An air compressor located under the table was used to provide two air channels for the experiment at a constant pressure of $2$ Bars. The first channel was aromatized by passing through a smell reservoir where citral in liquid form was stored. Both the aromatized and the control channel were controlled using the DMFCs which were connected through USB ports on the laptop.

The odour of citral was used as the delivered olfactory stimulus in this experimental study. This is a frequently encountered \changemarker{volatile organic compound (VOC)} reminiscent of the smell of lemon. Citral has a low detection threshold that ranges between $25$ to $350$ ppb~\cite{Citral16}. All the participants were able to detect it and no participant reported the smell as unpleasant. The range $I$ contained intensity stimulations that varied from faint up to more intense smell impulses without being irritating for the people who smelled them. The experimentation procedure was conducted under normal indoor lighting conditions in a room where air was constantly ventilated during the $4$ days of conducting the experiment. 

\subsubsection{Participants and Procedure}
Two groups of $10$ participants (11m, 9f, age range 19 to 35, $\mu$ = 31.6) from various academic and working disciplines. The average age of all the participants was $31.6$ years old. No participant reported olfactory deficits or temporary/permanent anosmia problems.

\changemarker{For each trial,} every participant had to smell the two bursts and then decide which one was the most intense by pressing one of the GUI buttons provided. During the tutorial session, pairs of $(C_{0}, C_{0})$ stimuli were presented and the participant was asked if the smell was familiar or not. All the $20$ participants were able to recognize the delivered odour. Participants were presented smells with the GUI providing context of the order and the participant could select the most intense stimulus from the GUI. Subsequently, the participant was permitted to select next, at their leisure, to continue onto the next trial. \changemarker{In the first phase each participant conducted eight trials and in the second they conducted ten trials.}

\begin{figure}[!h]
\centering
\includegraphics[scale=0.135]{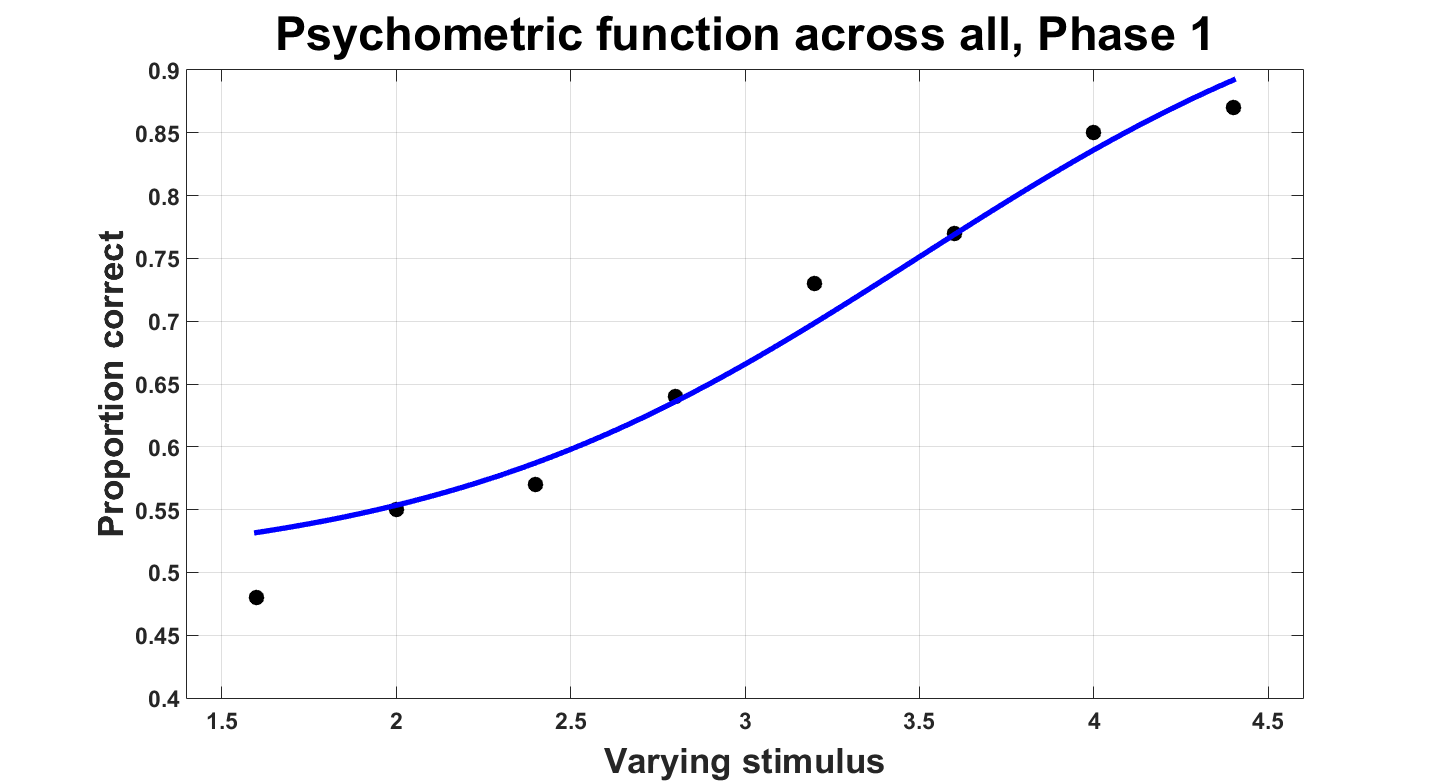}
\includegraphics[scale=0.135]{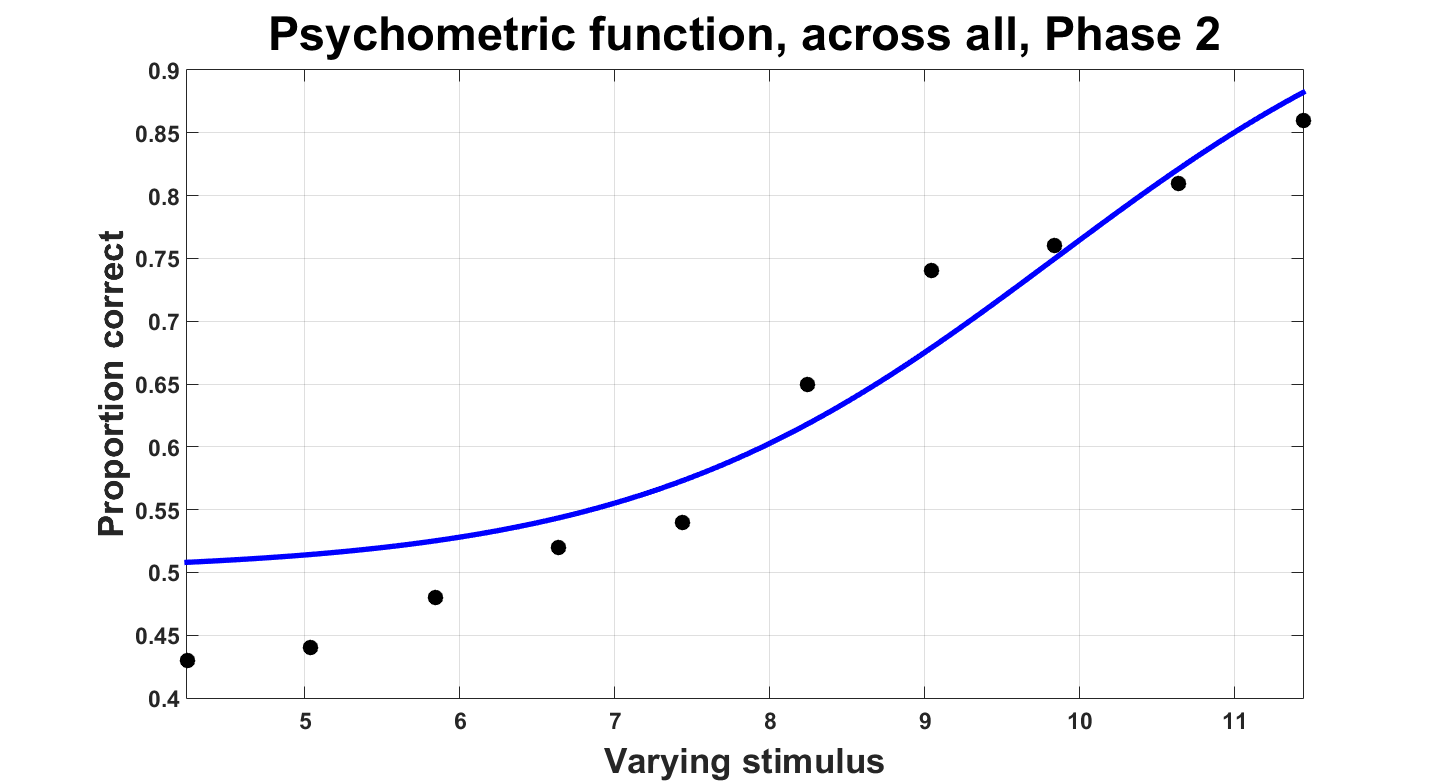}
\caption{Proportion of correct answers across all participants in Phase $1$ (Top) and Phase $2$ (Bottom). The blue colored curve is the Logistic Psychometric function fitted to the data in both cases.}
\label{fig:PFs}
\end{figure}

\subsubsection{JND Results}
\label{sec:JND_results}
Two JNDs were found in the operating range $I$ provided by the olfactory display at two different experimentation phases. The first estimated stimulus that elicited a perceptual difference had magnitude in the subinterval $I_{1}$ while the second was obtained using one step adjustment with the adaptive method described previously.

Phase $1$ started with a pedestal stimulus $C_{0} = 1.20$ ppm while there were $8$ varying stimuli to compare it with (see $I_{1}$ interval at Equation~\ref{eq:intervals}). Figure~\ref{fig:PFs} depicts the proportion of correct answers across all $10$ participants participated in this phase and the Logistic Psychometric function (PF) fitted to the data. This function was preferred over others as it most accurately fitted the obtained data using deviance criterion as a model comparison metric~\cite{KP09}. For the 2IFC method, the stimulus that corresponds to $75 \%$ of correct answers is given through the PF and corresponds to the magnitude \changemarker{$C_{1} = 3.50$ ppm}. Therefore, the first JND threshold is \changemarker{JND$_{1} = C_{1} - C_{0} = 2.30$ ppm.}

The slope of the PF expresses how participants' performance changes with increasing concentration magnitudes. The estimate for this value was \changemarker{$S_{1} = 0.14$ units of performance per ppm}. The standard error for the PF estimates for $C_{1}$ and $S_{1}$ were obtained based on a parametric bootstrap analysis for $N=1000$ generated data sets using the collected data of phase $1$. For all the generated data collections the parameters $C_{1}$ and $S_{1}$ were estimated and their standard deviation was computed. These were SD$_{C_{1}} = 0.12$ ppm and \changemarker{SD$_{S_{1}} = 0.25$ ppm.}

The goodness of fit for the selected PF was also computed to check if the selected PF sufficiently fits the data. The p-value for deviance found \changemarker{$P_{d} = 0.69 > 0.05$} meaning that the Logistic (\textbf{L}) PF is a representative statistical model for the given data. The same criterion with Weibul (\textbf{W}), Gumbel (\textbf{G}), Cumulative Normal (\textbf{CN}) and Hyperbolic Secant (\textbf{HS}) PFs resulted in an acceptable $P_{d} > 0.05$ but lower that the result given by the Logistic PF. The deviance values are given in Table~\ref{tbl:deviance}.

The Weber constant for the results of phase $1$ was \changemarker{$k = 1.91$}. Using the $C_{1}$ as the new pedestal, Weber's law (see formula~\ref{eq:Weber}) estimates that the new stimulus that elicits a perceptual difference is approximately $10.20$ ppm. Using this outcome the step was adjusted as $s = 0.8$ while a new experimental phase with a set of $10$ varying stimuli was initiated. This set is:

\begin{equation}
I_{a} = \{C_{1} + s, C_{1} + 2s, \cdots, 11.2\}
\end{equation}
The last varying stimulus was \changemarker{$11.50$ ppm} which is out of the olfactory display's operating range, therefore it was replaced with the interval's maximum. The proportion of correct answers and the PF function are given in Figure~\ref{fig:PFs}. The stimulus that elicited a perceptual difference was found to be \changemarker{$C_{1} = 10.13$ ppm} meaning that the second JND threshold is \changemarker{JND$_{2} = C_{2} - C_{1} = 6.63$ ppm}. This estimation is relatively close to Weber's law prediction and the small difference is attributed due to experimentation error. The slope of the new PF is \changemarker{$S_{2} = 0.07$ units of performance per ppm.} The standard error for both estimations is SD$_{C_{2}} = 0.23$ ppm and SD$_{S_{2}} = 0.13$ ppm using the same number of generated data sets as in phase $1$ for the parametric bootstrap. The deviance p-value for assessing goodness of fit was $P_{d} = 0.34 > 0.05$, higher than all the other PFs tested (see Table \ref{tbl:deviance}).

\begin{table}[!h]
\centering
\resizebox{.9\columnwidth}{!}{%
\begin{tabular}{lccccc}
\toprule
&\textbf{L}&\textbf{W}&\textbf{G}&\textbf{CN}&\textbf{HS}\\
\midrule
\textbf{Phase 1}&$0.6949$&$0.581$&$0.579$&$0.631$&$0.629$\\
\textbf{Phase 2}&$0.34$&$0.301$&$0.293$&$0.318$&$0.325$\\
\bottomrule
\end{tabular}%
}
\vspace{10pt}
\caption{Deviance p-values for the different statistical models tested for goodness of fit for each of the two experimental phases.}
\label{tbl:deviance}
\end{table}

\subsection{Discussion}
Smell transport simulation for different spatial discretisation levels and the JND experimental study yielded a number of potentially interesting findings. The experimental results clearly revealed that the human olfactory system (HOS) is relatively insensitive at small intensity variations. This result is confirmed from the high JND thresholds found at the whole operating range provided by the available hardware. The low standard deviations SD$_{C_{1}}$ and SD$_{C_{2}}$ of the estimates $C_{1}$ and $C_{2}$ that elicited perceptual differences show that the majority of the participants had difficulty in perceiving small intensity differences between the pedestal and varying stimuli. For olfactory stimulations of higher concentration the JND threshold is higher meaning that intensity variations are much more difficult to detect.

Although fine spatial discretisation levels of the computational domain can provide higher numerical accuracy in many applications, for olfaction, the JND threshold is clearly higher than the difference in odour intensity between the two extreme mesh sizes at every time point simulated. Accurate spatial discretisation of the computational domain in CFD simulations does not, on average, elicit quality improvements that can be consciously perceived by the users, hence the choice is made for olfaction to be a binary (on/off) choice in Experiment II.

\section{Experiment II: Tri-modal Resource Allocation}\label{sec:expFramework}
This section describes the details of this experiment including experimental layout, material preparation, procedure and participants.

\subsection{Design}
The experimental design extends the design of Doukakis et al. \cite{Doukakis2018}, whereby participants are asked to allocate a given budget of resources by adjusting the quality of the displayed visual and auditory stimuli, to now include the delivery of smell impulses in a binary (on/off) fashion. The reasoning of on/off smell was based on the results of Experiment I. It is well known that the operation of the human olfactory mechanism is not fully understood~\cite{Harel03, Nic09} and the evaluation of the incoming olfactory cues is highly subjective and based on previous experiences of the perceived scent~\cite{SSP04,Herz1996}. To avoid any bias of the quality levels due to these reasons, a two level scale that defines smell on or off is followed.

In this experiment, the audio-visual quality levels can be adjusted using the interlinked sliders of a GUI whilst two more controls are included for turning the smell stimulus on or off respectively. Physically-based simulations were used for the visual and aural stimuli whilst smell transport simulations were implemented for smell (see Section~\ref{sec:preexp}). The assigned budget allows the users to make audio-visual quality improvements and the cost of these improvements is deducted from the available budget. When the smell option is set to on, a percentage of the given budget is immediately reserved and the rest is given for audio-visual improvements.

The experiment was implemented with five distinct budgets across three different scenarios. One more scenario was used for training before the main experimental session. The experimental design is within participants and each participant was requested to make a judgement regarding the best perceived quality of all the possible combinations. The presentation order of the combinations was randomized to avoid any potential bias. In the rest of this section, the quality metrics for vision, audio and olfactory are explained. These metrics are used to derive quality levels for the three senses. The computation and metric selection for audio-visual stimuli was based on previous work \cite{Doukakis2018} and a short summary of these metrics is presented here for completeness.

\subsubsection{\changemarker{Visual}}
In the visual domain, quality is adjusted using image resolution. Resolution is a standardized metric and it can be abstracted from the underlying algorithm used for the image computation. Other factors that can adjust image quality (samples per pixel, textures, etc.) are kept constant to avoid an exponential growth of  the different possibilities. In this work, $240$ images were computed using path tracing \cite{Kaj86} starting from as low as $16 \times 9$ up to the highest resolution at Ultra HD (UHD) ($3840 \times 2160$). The normalized computational cost for all the images in this sequence can be given as:
\begin{equation}
\label{eq:_norm_graphics_cost}
C_{k}^{N_{V}} = \Big (\frac{k}{240} \Big )^2, \quad k = 1,2,\cdots,240.
\end{equation}
As the visual differences between successive images can be very small, the original sequence was filtered using the High Dynamic Range Visible Difference Predictor (HDR-VDP) model~\cite{narw14} to discard those pairs that elicit the same perceptual response to the average user. The final sequence included $80$ perceptually distinguishable images.

\subsubsection{\changemarker{Audio}}
In the auditory domain, quality was adjusted using the sampling frequency of the computed room impulse response (RIR). Again a ray tracing approach~\cite{Siltanen07} was used in the computation of the RIRs reaching a maximum sampling rate of $352,800$ Hz. Similarly to the visual domain, $240$ RIRs were computed at $240$ different sampling rates. Every RIR was convoluted with an anechoic sound stream that contained the audio context of every scenario. The normalized computational times of every RIR can be estimated as:
\begin{equation}
\label{eq:audio_cost}
C_{f_{k}}^{N_{A}} \approx \frac{f_{k}}{f_{240}}, \quad k = 1,2,\cdots,240,
\end{equation}
where $f_{k} < f_{240}$, and $f_{240} = 352800$ Hz the maximum sampling rate. Based on the frequency JND distribution in the auditory domain~\cite{Kollmeier2008}, a filtering of the RIRs gives $80$ log-spaced RIRs that clearly elicit sound differences after the convolution with the anechoic stram.

\subsubsection{Olfactory}
In the olfactory domain, a similar technique was followed for defining normalized costs for the olfactory cues. This technique is based on the perceptual properties found in Experiment I. The theoretical cost for an olfactory cue is estimated using the concentration results over virtual time found during the smell transport simulations. Specifically, for a given scenario, denoted by $S$, and a mesh size, denoted by $M$, the notation $VT_{M}^{S}$ denotes the virtual time moment when the simulation reaches the concentration $C_{1} = 3.49$ ppm at the probe location for the chosen mesh and scenario. The concentration level $C_{1}$ is the first olfactory stimulus that elicits a perceptual difference starting from concentration $C_{0} = 1.20$ ppm and was preferred as it is a medium intensity olfactory stimulus without being too faint or strong when sniffed. Using the notation $PT_{M}^{S}$ for denoting the physical computation time needed for computing all the intermediate virtual times including $VT_{M}^{S}$, the normalized cost for smell is defined as:
\begin{equation}
A_{M}^{S} = \frac{PT_{M}^{S}}{PT_{1M}^{S}},
\label{eq:smell_norm_cost}
\end{equation}
where the subscript ``$1$M'' indicates the highest mesh quality used in the smell transport simulations ($1$ million CVs). The theoretical cost $A_{M}^{S}$ is always between $[0,1]$ while it is an increasing function of the mesh size, i.e. $A_{M}^{S_{1}} \leq A_{M}^{S_{2}}$ for meshes $S_{1}$, $S_{2}$ where $S_{1}$ has less CVs than $S_{2}$.

\subsubsection{Tri-modal cost interactions}
The choice of normalized cost functions in the audio, visual and olfactory domains, allows to investigate which of the three senses is more important for a given budget size without considering the problem of large differences in physical computation time for rendering an image,  RIR or simulating an odour impulse in the VE. Also, normalisation of the costs makes the experimental results independent of the algorithmic strategy selected for computing the three stimuli. In this study, the total budget size is always distributed among the costs of a visual, auditory and an olfactory stimulus (when on). The budget sizes used in this study are given in Table \ref{budget_table}. \changemarker{Note that $\mathbf{B_{5}}$ has fewer levels as the lowest qualities would not be available for this budget since the overall budget is higher without increasing the maximum qualities available for both audio and visuals.}

\begin{table}[!h]
\caption{Theoretical budgets used in this experimental study along with their notation letters and the number of quality levels available when a budget is applied.}
\label{budget_table}
\centering
\renewcommand{\arraystretch}{1.1}
\scalebox{0.9}{
\begin{tabular}{lccccc}
\toprule
\textbf{Budget letter} &  $\mathbf{B_{1}}$ & $\mathbf{B_{2}}$ & $\mathbf{B_{3}}$ & $\mathbf{B_{4}}$ & $\mathbf{B_{5}}$\\
\midrule
\textbf{Budget} & $0.0625$ & $0.11$ & $0.25$ & $1$ & $1.12$\\
\textbf{Total Number of Levels} & $28$ & $38$ & $48$ & $80$ & $48$\\
\bottomrule
\end{tabular}%
}
\end{table}

The wide range of budget sizes ($B_{1}$-$B_{5}$) allows to investigate users' allocation preferences both with and without budget constraints. Using equation~\ref{eq:smell_norm_cost}, and $M = 1K$, the normalized costs for the smell impulse can be computed for all four scenarios of interest. The theoretical cost for smell was estimated using the physical computation times obtained from the coarsest mesh size, therefore, the selection of $1K$ is justified from the findings of Experiment I, where the selection of the mesh size for simulating smell transport has no perceptual effect to the user.

\begin{table}[!h]
\caption{Theoretical costs for smell across the four scenarios selected. The first row shows the virtual time (sec) required to reach concentration $C_{1}$. The second row shows physical computation times (sec) for the coarsest mesh (1K). The third row shows physical computation times (sec) for the finest mesh (1M). The last row gives theoretical costs for smell as ratios of the second and third row.}
\label{smell_costs_table}
\centering
\resizebox{\columnwidth}{!}{%
\renewcommand{\arraystretch}{1.1}
\scalebox{0.85}{
\begin{tabular}{lrrrr}
\toprule
\textbf{Scenario} & Bathroom & Car & Kitchen & Kitti \\
\midrule
$VT_{1K}^{S}$ & $646$ & $458$ & $1685$ & $1199$\\ 							
$PT_{1K}^{S}$ &$229.2$  & $123.6$  &$718.8$ &$481.3$ \\					
$PT_{1M}^{S}$ &$6222.5$ & $4164.7$ & $17900.3$ & $15228.9$\\		
$A_{1K}^{S} $ & $0.037$  & $0.029$ & $0.040$ & $0.032$\\					
\bottomrule
\end{tabular}%
}
}
\end{table}


\subsection{Materials}
\begin{figure}[!t]
\centering
\includegraphics[scale=0.2]{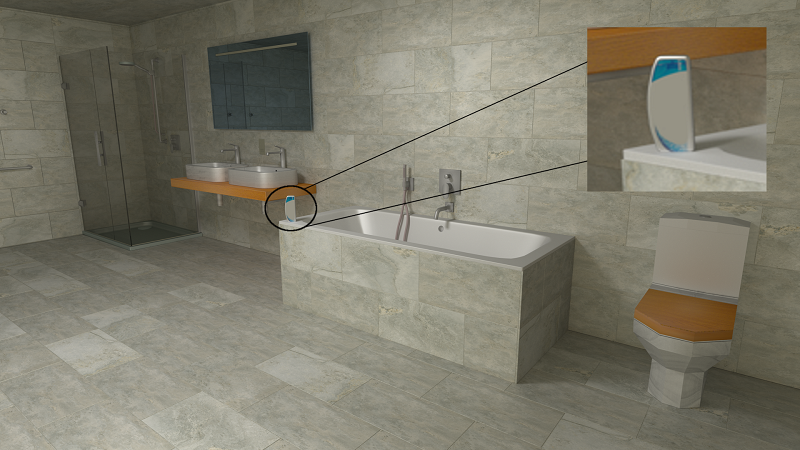}
\includegraphics[scale=0.2]{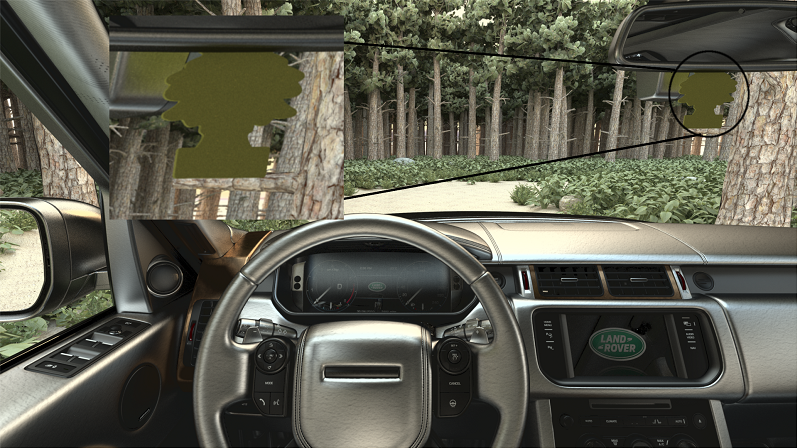}\\
\vspace{1mm}
\includegraphics[scale=0.2]{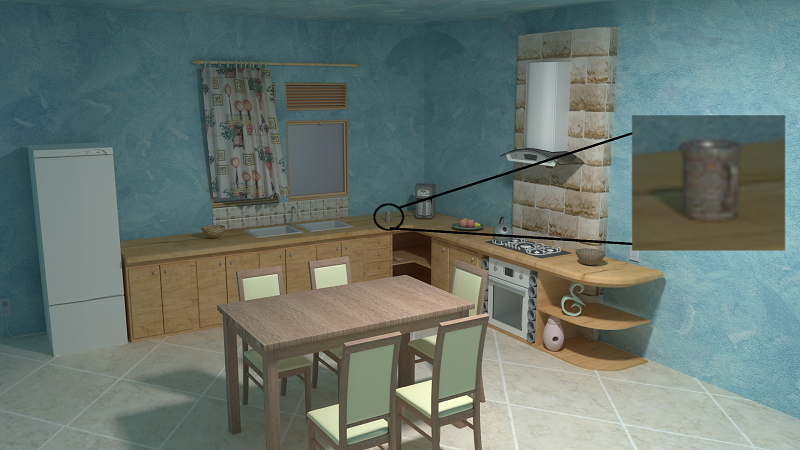}
\includegraphics[scale=0.2]{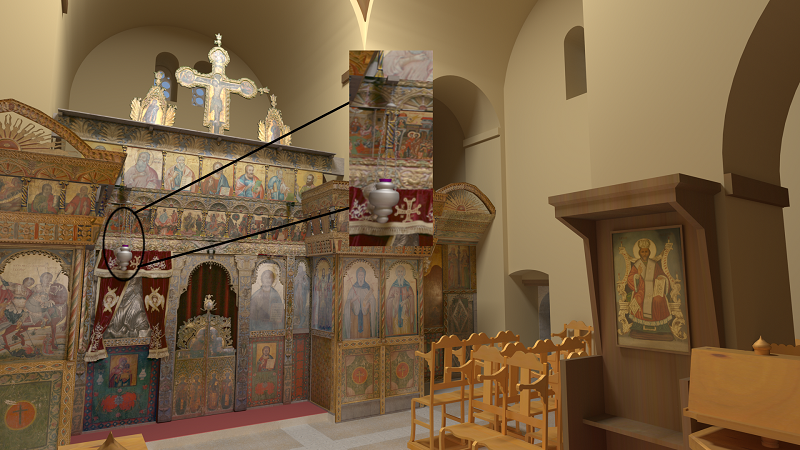}\\
\caption{Scenarios of this experimental study. From Left to Right and Top to Bottom: Bathroom, Car, Kitchen, Kitti. The circular areas show the smell source.}
\label{fig:VE_scenarios}
\end{figure}

\begin{figure}[!t]
\centering
\includegraphics[scale=0.2]{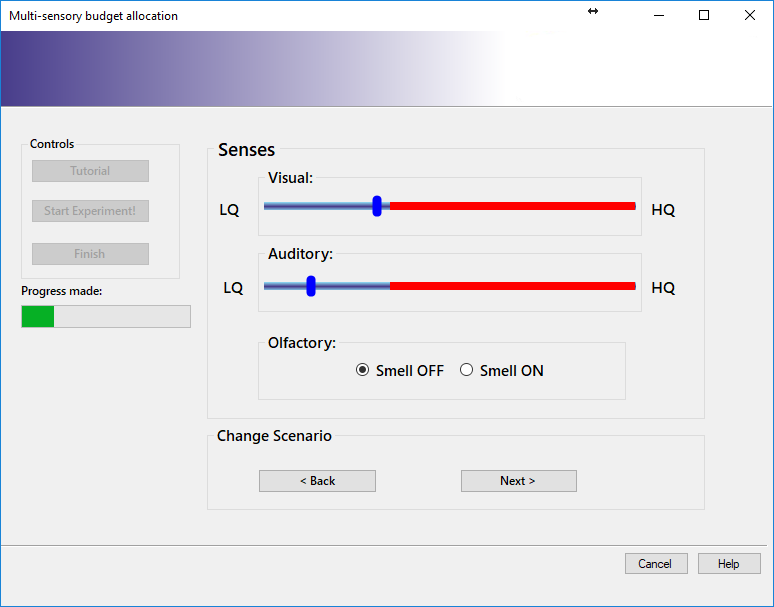}
\includegraphics[scale=0.2]{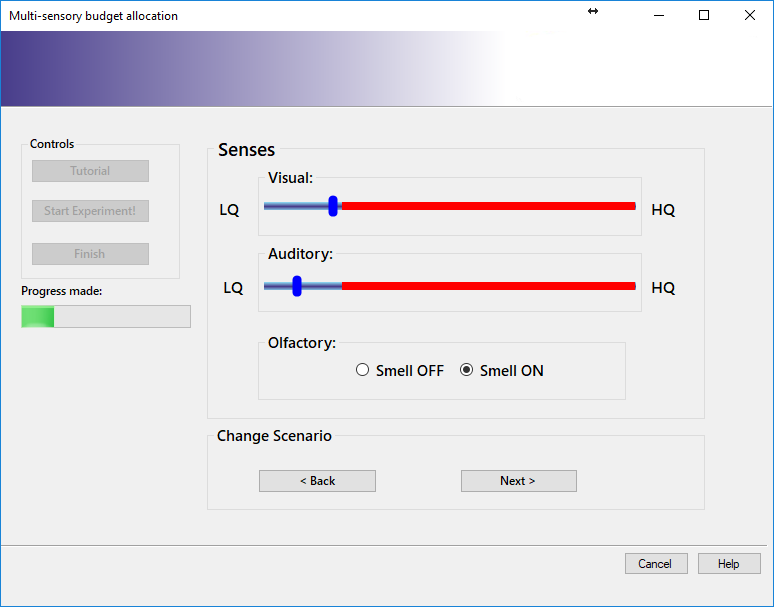}
\caption{Snapshots of the software used in the experimental study. Left: An instance of the budget size $B_{2}$ when the ``Smell OFF'' control is activated. Right: An instance of the budget size $B_{2}$ when the ``Smell ON'' control is activated.}
\label{fig:exp_GUI}
\end{figure}

For this study, two monitors (First: 28$^{\prime\prime}$ Samsung U28D590 ultra HD LED monitor used for visual stimuli, Second: Dell UltraSharp $19^{\prime\prime}$ used for the GUI) and a set of headphones (Sennheiser HD 380 pro) were used for conducting this experiment. In addition, the olfactory display used of the preliminary experimental study was used for the delivery of the smell impulses. All the experimental trials were conducted in a silent room with constant ventilation.
The same four scenarios used for smell transport simulations in Experiment I were used. The smell of citral, was congruent to all of them. Figure~\ref{fig:VE_scenarios} depicts the experimental scenarios and zooms in the source of smell used for the simulation of odour transport. From the user's perspective, at least one object could have been the source of smell and no information was given to the participants regarding the source of the odour. A GUI application was used for carrying out the experimental study, see Figure~\ref{fig:exp_GUI}.

The user becomes aware of the constraint in the budget as the red colored stripes increase in length and the number of available levels for audio-visual quality improvements are reduced. When the control for smell was turned on, a smell impulse of $1$ second was delivered to the user at close vicinity to the nose. The delivery of smell bursts instead of a continuous flow was preferred so as to avoid user's fast adaptation to the released odour. A control channel was also used to release air bursts of $1,000$ ml/min flow rate and convect the smell bursts away from participant's location. Five air burst of frequency $400$ ms were released when the user was pressing the ``Next''  or the ``Smell OFF'' buttons to go to the next trial or choose not to allocate resources when the ``Smell ON'' was enabled respectively. The delivery of the audio and visual stimuli was delayed by $240$ ms when the smell was set on in the experiment to compensate the olfactory display's onset time. The objective was to deliver all three sensory stimuli simultaneously without the participants perceiving temporal delays.

The use of the air channel was very important as it didn't allow adaptation of the user to the smell cue. The air channel intended to disperse the smell plume quickly at the ventilation system which was right behind and above participant's head. Also in situations where participants wanted to try the smell cue and disable it immediately, the air channel assisted to remove the residues of the smell out immediately in an attempt to disable the sensory cue as quickly as possible.

At the beginning of each experimental trial the two sliders are positioned at the start of the slider bars which correspond to a ``null'' stimulus. The ``null'' stimulus configuration includes the delivery of a grey image and a silent track while the smell control is pre-selected to be ``OFF''. The initial selection of the grey image aims to neutralize participants' eyes and is suggested by~\cite{ITUStandard2}. When budget $B_{5}$ is available for resource allocation, the audio and visual thumbs start from a medium quality level that has cost equal to $B_{5} - B_{4}$. The initial configuration of the thumbs allows the participants to explore all the available quality levels before deciding which quality level is desired. The alternative option to start the thumbs at random positions was not preferred to avoid biasing the participant with a thumb configuration which does not represent his/her actual preferences. At the beginning of each trial, the two thumbs are independent of each other until the sum of the theoretical costs for the three sensory stimuli exceeds the budget given for the trial. After the first attempt to exceed the budget, the thumb controls become dependent. At any point, if the user activates the smell impulse, the GUI sliders adjust to the new lower budget.


The transition from independent to dependent slider controls for adjusting audio-visual quality changes to either the senses of vision or hearing the quality of the other stimulus so as the budget to remain always constant and equal to the current budget size. The addition of the olfactory impulses does not disrupt this mechanism as the cost for a smell impulse is immediately allocated after the control is turned on.

\subsection{Participants and Procedure}
A total of $25$ participants ($13$m, $12$f, age range: $23$ to $46$, $\mu = 36.2$) volunteered.
All of them had normal or corrected to normal vision. None of them reported hearing or temporary/permanent smell problems. All the participants recruited for this experiment had no participation in Experiment I or were aware of its purpose.

Every experimental session was initiated with a tutorial where the participants had the chance to familiarize with the task of allocating resources. The Kitti scenario was used for training
with all possible budgets. During the tutorial, the experimenter had the chance to explain the objectives and purpose of the study to the participants for about $15$ minutes before the main session. The participants were asked to judge and form the best multi-sensory experience using the controls of the GUI.


\subsection{Results}\label{sec:results}
The analysis of the results is divided into two parts for better clarity. The first part includes an ANOVA via a $3$ (\textit{scenario}) $\times$ $5$ (\textit{budget}) factorial design for studying the effect of the independent variables on participants' decision to enable the smell impulse on and the second part describes a multivariate analysis of variance (MANOVA) for examining the effect of the two independent variables on the visual and auditory allocation preferences.

\subsubsection{Analysis of the smell preferences}
\begin{figure}[!h]
\centering
\includegraphics[scale=0.23]{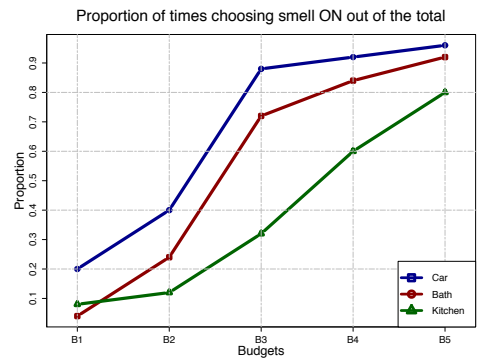}
\caption{Proportion of times smell stimulus was set on in each of the three scenarios and across all different budgets.}
\label{fig:smell_prop_ON}
\end{figure}

Figure~\ref{fig:smell_prop_ON} depicts the proportion of times that people preferred to receive a smell impulse along with an audio/visual stimulus for each of the three scenarios and across every budget. At small budget sizes, participants choose not to allocate resources for enabilng the olfactory stimuli. As the budget increases, the frequency people select to receive an olfactory stimulus increases significantly. Furthermore, people tend to enable the control for smell more often in the Car and Bathroom VEs compared to the Kitchen VE. Participants' preference to these scenarios might be explained due to the ease of visually locating the smell source in the scene.

The main effect of budget was significant with \changemarker{$F(4,96) = 61.47$}, $p < 0.05$ indicating a difference in the proportions the subjects selected to receive a smell impulse across the five budgets. The main effect of budget did not violate the assumption of sphericity (via Maulchy's test, $p > 0.05$). The main effect of scenario was also significant \changemarker{$F(2,48) = 18.42$}, $p < 0.05$ indicating a difference in the proportions the smell burst was received by the participants across the three scenarios. The main effect for scenario did not also violate the assumption of sphericity. The interaction of budget $\times$ scenario was also examined and it was not found to be statistically significant, \changemarker{$F(8,192) = 1.76$}, $p>0.05$.

Contrast comparisons for proportions were conducted between groups of budgets using post-hoc proportion tests with Bonferroni corrections. These results are presented in Table~\ref{tbl:contrast_between_budgets}. The results show that small budget sizes ($B_{1}, B_{2}$) are not significantly different indicating that the budget size is still not sufficient for distributing resources to smell. Large budget sizes are also not significantly different ($B_{4},B_{5}$) meaning that people tend to enable the smell stimulus approximately the same number of times during their allocation.

\begin{table}[!h]
\centering
\resizebox{0.35\textwidth}{!}{%
\begin{tabular}{lllllll}
\toprule
\textbf{Scenario}&  \multicolumn{5}{c}{\textbf{Budget size}} &$\mathbf{p-}$\textbf{value} \\
\midrule
Car       	&\mbl{c1}$B_{1}$&$B_{2}$\mbr{c1}{red}&\mbl{c2}$B_{3}$&$B_{4}$&$B_5$\mbr{c2}{blue}& $< 0.05$ \\
Bath        &\mbl{c1}$B_{1}$&$B_{2}$\mbr{c1}{red}&\mbl{c2}$B_{3}$&$B_{4}$&$B_5$\mbr{c2}{blue}& $< 0.05$ \\
Kitchen   &\mbl{c3}$B_{1}$&$B_{2}$&\mbl{c4}$B_{3}$\mbr{c3}{red}&\mbl{c5}$B_{4}$\mbr{c4}{green}&$B_5$\mbr{c5}{blue}& $< 0.05$ \\
\midrule
All			&\mbl{c6}$B_1$&$B_2$\mbr{c6}{red}&\mbl{c8}$B_3$\mbr{c8}{green}&\mbl{c7}$B_4$&$B_5$\mbr{c7}{blue}& $< 0.05$ \\
\bottomrule
\end{tabular}%
}
\caption{Contrast comparisons for proportions of smell between budgets. Proportions of budgets with no significant differences are grouped together.}
\label{tbl:contrast_between_budgets}
\end{table}

Contrast comparisons for proportions were also conducted between groups of scenarios using post-hoc proportion tests and applying Bonferroni corrections.
For $B_1$ and $B_2$ no significant changes were found. For the remaining three budget sizes, the group $\{$Car, Bath$\}$ is significantly different from the group $\{$Kitchen$\}$ indicating that participants clearly prefer to enable the smell on more frequently at scenarios of the former group.


\subsubsection{Analysis on the audio-visual preferences}
\begin{figure}[!h]
\centering
\includegraphics[scale=0.22]{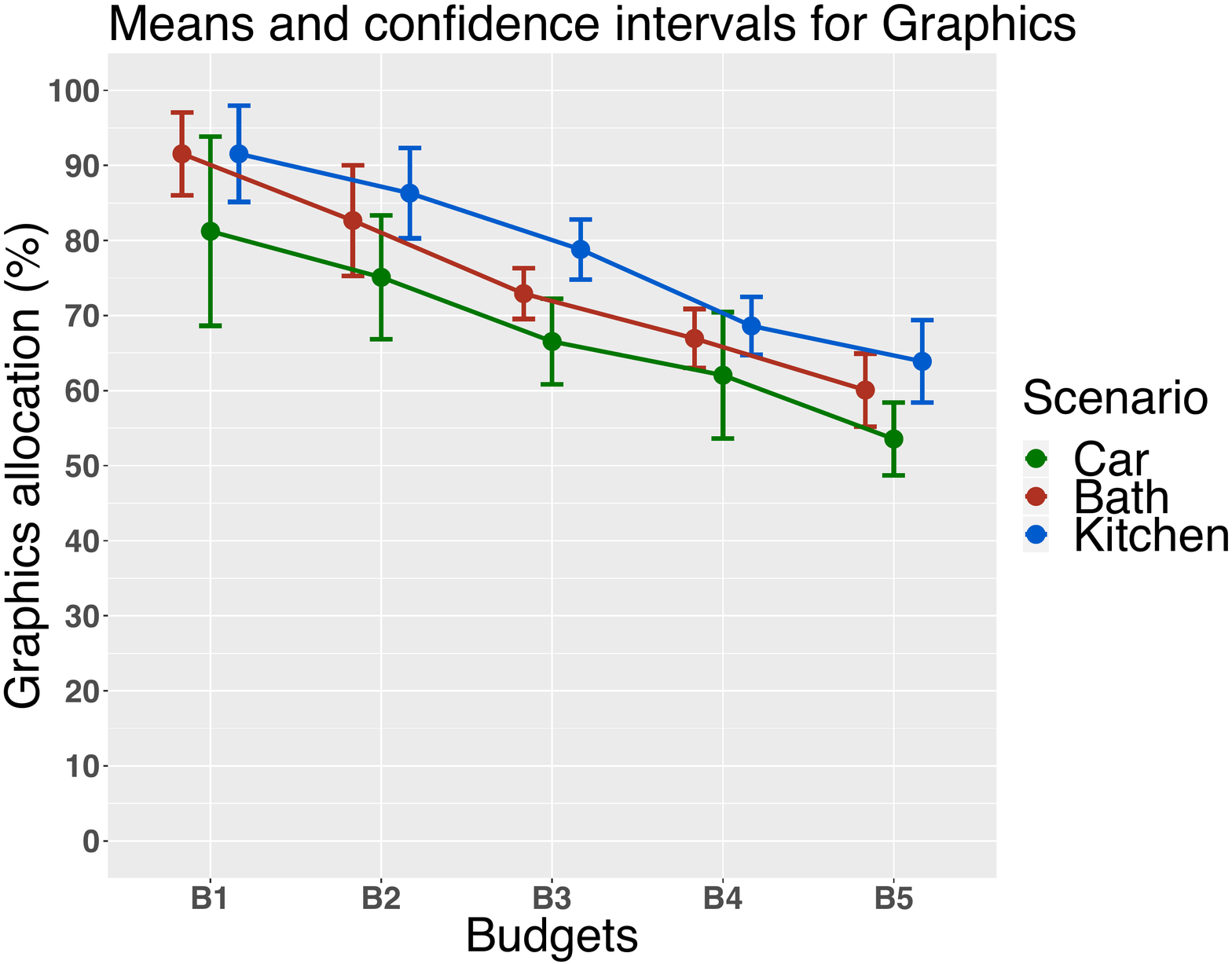}
\includegraphics[scale=0.22]{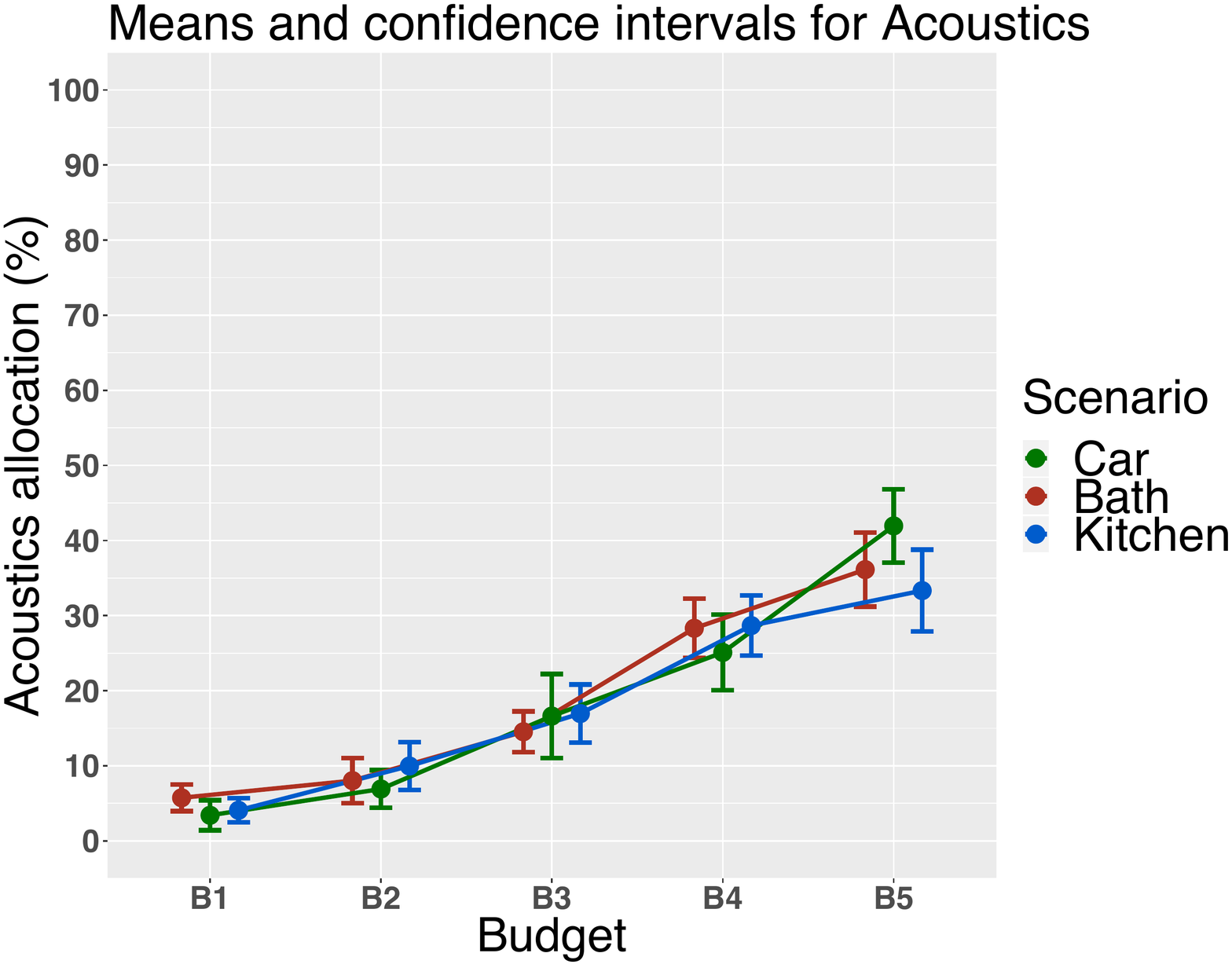}
\caption{Mean allocation percentages and confidence intervals for every scenario and across all the budget sizes Top: Graphics, Bottom: Acoustics. Jittering has been applied to the two plots for better visualisation.}
\label{fig:audio_visual_percentages}
\end{figure}

Figure~\ref{fig:audio_visual_percentages} depicts the mean percentages and confidence intervals for graphics and acoustics quality for each scenario and budget.
As can be seen, the visual and auditory percentages are negatively correlated. For small budget sizes participants prioritise visuals and allocate the majority of the budget. As the budget size increases a balanced distribution of resources is preferred as the available amount permits it.

MANOVA analysis was applied to examine the effect of the budget and scenario on the audio-visual allocation percentages. For the acoustic quality percentage, the main effect of budget was significant with $F(4,96) = 153.85$, $p < 0.05$ while the effect of scenario was not found to be statistically significant with $F(2,48) = 0.0241$, $p > 0.05$. The main effect of budget and scenario did not violate the assumption of sphericity (via Maulchy's test, $p > 0.05$). The interaction scenario $\times$ budget was not found significant, $F(8,196) = 0.012$ $p > 0.05$.

For the visual quality percentage, the main effect of budget was significant with \changemarker{$F(4,96) = 42.11$}, $p < 0.05$. The effect of scenario was also found to be statistically significant for the graphics allocation percentage with \changemarker{$F(2,48) = 11.03$}, $p < 0.05$. The main effect of both factors did not violate the assumption of sphericity (via Maulchy's test, $p > 0.05$). The interaction scenario $\times$ budget was not found significant, \changemarker{$F(8,196) = 0.94$}, $p > 0.05$.

Contrast comparisons using post-hoc t-tests were also conducted to investigate groups of budgets that are not significantly different across the scenarios and also for finding groups of scenarios that are not significantly different across budgets. These contrasts are given in Table~\ref{tbl:graphicsTtests} and Table~\ref{tbl:acousticsTtests} for graphics and acoustics respectively.

\begin{table}[!ht]
\parbox{.49\linewidth}{
\centering
\resizebox{0.52\columnwidth}{!}{%
\begin{tabular}{lllllll}
\toprule
\textbf{Scenario}&  \multicolumn{5}{c}{\textbf{Budget size}} &$\mathbf{p-}$\textbf{value} \\
\midrule
Car       &\mbl{c1}$B_{1}$&$B_{2}$&$B_{3}$&$B_{4}$\mbr{c1}{red}&$B_5$& $< 0.05$ \\
Bath       &\mbl{c1}$B_{1}$&$B_{2}$\mbr{c1}{red}&\mbl{c2}$B_{3}$&\mbl{c3}$B_{4}$\mbr{c2}{blue}&$B_5$\mbr{c3}{green}& $< 0.05$ \\
Kitchen   &\mbl{c4}$B_{1}$&$B_{2}$\mbr{c4}{red}&$B_{3}$&\mbl{c5}$B_{4}$&$B_5$\mbr{c5}{green}& $< 0.05$ \\
\midrule
All&\mbl{c6}$B_1$&$B_2$\mbr{c6}{red}&$B_3$&\mbl{c7}$B_4$&$B_5$\mbr{c7}{blue}& $< 0.05$ \\
\bottomrule
\end{tabular}%
}
}
\hfill
\parbox{.49\linewidth}{
\centering
\resizebox{0.52\columnwidth}{!}{%
\begin{tabular}{lllll}
\toprule
\textbf{Budget}&  \multicolumn{3}{c}{\textbf{Scenarios}} &$\mathbf{p-}$\textbf{value} \\
\midrule
$B_1$     &\mbl{c1}Car&Bath&Kitchen\mbr{c1}{red}& $< 0.05$ \\
$B_2$     &\mbl{c2}Car&\mbl{c3}Bath\mbr{c2}{red}&Kitchen\mbr{c3}{blue}& $< 0.05$ \\
$B_3$     &\mbl{c4}Car&\mbl{c5}Bath\mbr{c4}{red}&Kitchen\mbr{c5}{blue}& $< 0.05$ \\
$B_4$     &\mbl{c6}Car&Bath&Kitchen\mbr{c6}{red}& $< 0.05$ \\
$B_5$     &\mbl{c7}Car&\mbl{c8}Bath\mbr{c7}{red}&Kitchen\mbr{c8}{blue}& $< 0.05$ \\
\midrule
 All      		&Car&\mbl{c9}Bath&Kitchen\mbr{c9}{red}& $< 0.05$ \\
\bottomrule
\end{tabular}%
}
}
\caption{Left: Contrast comparisons between budgets at every scenario and across all the scenarios for graphics. Right: Contrast comparisons between scenarios at every budget and across all the budgets for graphics. Budgets or scenarios with no significant difference are grouped together.}
\label{tbl:graphicsTtests}
\end{table}

As can be seen from Table~\ref{tbl:graphicsTtests}, small budget sizes ($B_{1},B_{2}$) are not significantly different meaning that participants follow similar allocation strategy for graphics for these budgets independently of the scenario. The same argument also holds for very large budget sizes ($B_{4},B_{5}$). As far as the scenarios are concerned, subjects tend to increase graphics quality in a similar way for the Kitchen and Bath scenarios.

\begin{table}[!ht]
\parbox{.49\linewidth}{
\centering
\resizebox{0.52\columnwidth}{!}{%
\begin{tabular}{lllllll}
\toprule
\textbf{Scenario}&  \multicolumn{5}{c}{\textbf{Budget size}} &$\mathbf{p-}$\textbf{value} \\
\midrule
Car       &\mbl{c1}$B_{1}$&$B_{2}$\mbr{c1}{red}&\mbl{c2}$B_{3}$&$B_{4}$\mbr{c2}{blue}&$B_5$& $< 0.05$ \\
Bath       &\mbl{c3}$B_{1}$&$B_{2}$\mbr{c3}{red}&$B_{3}$&\mbl{c4}$B_{4}$&$B_5$\mbr{c4}{blue}& $< 0.05$ \\
Kitchen   &$B_{1}$&$B_{2}$&$B_{3}$&\mbl{c5}$B_{4}$&$B_5$\mbr{c5}{blue}& $< 0.05$ \\
\midrule
All&\mbl{c6}$B_1$&$B_2$\mbr{c6}{red}&$B_3$&\mbl{c7}$B_4$&$B_5$\mbr{c7}{blue}& $< 0.05$ \\
\bottomrule
\end{tabular}%
}
}
\hfill
\parbox{.49\linewidth}{
\centering
\resizebox{0.52\columnwidth}{!}{%
\begin{tabular}{lllll}
\toprule
\textbf{Budget}&  \multicolumn{3}{c}{\textbf{Scenarios}} &$\mathbf{p-}$\textbf{value} \\
\midrule
$B_1$     &\mbl{c1}Car&\mbl{c2}Kitchen\mbr{c1}{red}&Bath\mbr{c2}{blue}& $< 0.05$ \\
$B_2$     &\mbl{c3}Car&Kitchen&Bath\mbr{c3}{red}& $< 0.05$ \\
$B_3$     &\mbl{c4}Car&Kitchen&Bath\mbr{c4}{red}& $< 0.05$ \\
$B_4$     &\mbl{c5}Car&Kitchen&Bath\mbr{c5}{red}& $< 0.05$ \\
$B_5$     &\mbl{c6}Car&Kitchen&Bath\mbr{c6}{red}& $< 0.05$ \\
\midrule
 All      		&\mbl{c7}Car&Kitchen&Bath\mbr{c7}{red}& $< 0.05$ \\
\bottomrule
\end{tabular}%
}
}
\caption{Left: Contrast comparisons between budgets at every scenario and across all the scenarios for acoustics, Right: Contrast comparisons between scenarios at every budget and across all the budgets for acoustics. Budgets or scenarios with no significant difference are grouped together.}
\label{tbl:acousticsTtests}
\end{table}

Table~\ref{tbl:acousticsTtests} shows that none of the scenarios is significantly different from the others when distributing resources for aural quality improvements. As far as the budgets are concerned, the same trend for graphics holds also for the acoustics for the effect of the budget size. 


\section{Estimation model}\label{sec:models}
The experimental data was used in the construction of a statistical model which aims to estimate three quantities given a budget, and optionally the scenario. The first quantity is the probability that expresses whether smell should be given for an input budget and scenario while the other two estimations are proportions of the total budget that should be devoted to visual and aural quality.

As the decision to turn the smell stimulus on or off is based on a categorical variable, a logistic regression was used to model it. Specifically, if the probability to give a smell impulse is denoted as $p$ then the ratio $\frac{p}{1-p}$ expresses the odds of enabling the smell impulse release. This smell on/off decision can be modeled as:
\begin{equation}
\log \Big (\frac{p}{1-p} \Big) = \hat{\beta_{i}^{s}} + \hat{\beta_{b}^{s}} \cdot \text{budget} + \mathbbm{1}_{\text{S}} \cdot \hat{\gamma_{S}^{s}},
\label{eq:M1smell}
\end{equation}
where the coefficients $\hat{\beta_{i}^{s}}$, $\hat{\beta_{b}^{s}}$ and $\hat{\gamma_{S}^{s}}$ are the least squares regression estimates. The subscripts $i$ and $b$ are shorthand for the intercept and the budget respectively, and the subscript $S$ is shorthand for scenario.
The motivation for scenario specificity is dictated by the significant differences identified between the scenarios in the experimental data.
$\hat{\gamma_{S}^{s}}$ acts as a scenario specific offset to the intercept which can improve the model for each scenario. The indicator function $\mathbbm{1}_{\text{S}}$ denotes whether the scenario is included as part of the model.

For an input budget size, if the signum of the right hand side of Equation~\ref{eq:M1smell} is positive then it can be inferred that $p > 1-p$ and thus a smell impulse should be delivered to the user. The above model is used to estimate whether the smell should be turned on/off and is only one component of the whole model. The full model is also composed of audio-visual percentage allocation components and can be written in matrix form as:

\begin{equation}
\left[
\begin{tabular}{c}
$\hat{S_{1}}$\\
$\hat{V_{1}}$\\
$\hat{A_{1}}$
\end{tabular}
\right]=
\left[
\begin{tabular}{ccc}
$\hat{\beta_{i}^{s}}$ &  $\hat{\beta_{b}^{s}}$ & $\hat{\gamma_{\text{S}}^{s}}$ \\
$\hat{\beta_{i}^{v}}$ &  $\hat{\beta_{b}^{v}}$ & $\hat{\gamma_{\text{S}}^{v}}$ \\
$\hat{\beta_{i}^{a}}$ &  $\hat{\beta_{b}^{a}}$ & $\hat{\gamma_{\text{S}}^{a}}$
\end{tabular}
\right] \cdot
\left[
\begin{tabular}{c}
$1$ \\
$\text{budget}$ \\
$\mathbbm{1}_{\text{S}}$
\end{tabular}
\right],
\label{eq:M1_model}
\end{equation}
where the letters $\hat{V_{1}}$ and $\hat{A_{1}}$ are the vision and audio allocation estimations while $\hat{S_{1}} = \log \Big( \frac{p}{1-p} \Big)$ is the model component for smell as given in formula~\ref{eq:M1smell}.

The remainder of the paper will consider two forms of the model for both testing and evaluation: $\mathbf{M_{1}}$ refers to $\mathbbm{1}_{\text{S}} = 0$ where the scenario is not considered, and $\mathbf{M_{2}}$ refers to $\mathbbm{1}_{\text{S}} = 1$ which considers the scenario in the model.

For $\mathbf{M_{1}}$, hypothesis tests of the form:
\begin{equation*}
H_{0}: \beta_{z}^{k} = 0  \text{ vs }H_{a}: \beta_{z}^{k} \neq 0, \quad z \in \{i,b\} \text{  and  } k \in \{s,v,a\}
\end{equation*}
were conducted to check the significance of the unknown parameters $\beta_{i}^{s}, \beta_{b}^{s}, \beta_{i}^{v}, \beta_{b}^{v}, \beta_{i}^{a}, \beta_{b}^{a}$. In all these tests the null hypothesis is rejected indicating that the respective least squares regression estimators ($\hat{\beta_{i}^{s}}, \hat{\beta_{b}^{s}}, \hat{\beta_{i}^{v}}, \hat{\beta_{b}^{v}}, \hat{\beta_{i}^{a}}, \hat{\beta_{b}^{a}}$) should remain in the model formulation. The six estimates on the right hand side of the above equation are given in Table~\ref{tbl:M1_reg_coefficients}.

\begin{table}[!h]
\centering
\resizebox{0.4\textwidth}{!}{%
\begin{tabular}{lcccccc}
\toprule
Model&$\hat{\beta_{i}^{s}}$ & $\hat{\beta_{b}^{s}}$ & $\hat{\beta_{i}^{v}}$  & $\hat{\beta_{b}^{v}}$ & $\hat{\beta_{i}^{a}}$ & $\hat{\beta_{b}^{a}}$\\
\midrule
$\mathbf{M_{1}}$&$-1.31$&$0.03$&$84.89$&$-0.25$&$4.72$&$0.32$\\
\bottomrule
\end{tabular}%
}
\caption{Least squares regression estimates of the multivariate model $\mathbf{M_{1}}$. The subscripts ``i'' and ``b'' are used as shorthands for intercept and budget respectively. The superscipts ``s'', ``v'', ``a'', are used instead of smell, vision and audio respectively.}
\label{tbl:M1_reg_coefficients}
\end{table}

The coefficients of determination $R^{2}$ and $R^{2}_{\text{adj}}$ for the goodness of fit for model $\mathbf{M_{1}}$ are given in Table~\ref{tbl:M1_deter_coefficients}. The superscript in parenthesis is used to denote the sense. As the smell estimation is based on logistic regression the first coefficient of determination is based on the Akaike Information Criterion (AIC) (the lower the better) while the other two for graphics and audio are the Nagelkerke coefficients of determination (the higher the better).

\begin{table}[!h]
\centering
\resizebox{0.4\textwidth}{!}{%
\begin{tabular}{lrrrrr}
\toprule
Model &$R^{2(s)}$ & $R^{2(v)}$ & $R^{2(v)}_{\text{adj}}$& $R^{2(a)}$ & $R^{2(a)}_{\text{adj}}$ \\
\midrule
$\mathbf{M_{1}}$ & $24.80$ & $0.37$ & $0.36$ & $0.61$ & $0.61 $ \\
$\mathbf{M_{2}}$ & $18.40$ & $0.40$ & $0.40$ & $0.65$ & $0.62 $ \\
\bottomrule
\end{tabular}%
}
\caption{Coefficients of determination for the multivariate models $\mathbf{M_{1}}$ and $\mathbf{M_{2}}$. The \changemarker{Akaike information criterion} was used for the $R^{2(s)}$. The Nagelkerke coefficient is used for vision and audio.}
\label{tbl:M1_deter_coefficients}
\end{table}

$\mathbf{M_{2}}$ considers the scenario as input to the model, and therefore can better match the experimental results than $\mathbf{M_{1}}$ which is based an the average across all scenarios. $\mathbf{M_{2}}$ introduces extra parameters $\gamma_{S}^{s}$, $\gamma_{S}^{v}$ and $\gamma_{S}^{a}$ which are examined to see whether they are statistically significant and should be kept in the model $\mathbf{M_{2}}$. Hypothesis tests of the following form:

\begin{equation}
H_{0}: \gamma_{S}^{k} = 0  \text{ vs }H_{a}: \gamma_{S}^{k} \neq 0,  \quad k \in \{s,v,a\},
\label{eq:hyp_test}
\end{equation}
were conducted to see the significance of these parameters in the Kitchen and Car scenario. For the Kitchen scenario ($S = K$), it was found that for the parameter $\gamma_{K}^{v}$ the $H_{0}$ is not rejected ($p > 0.05$) and the same also holds for the parameter $\gamma_{K}^{a}$. For the car scenario, $\gamma_{C}^{a}$ is also not rejected. The latter parameters were expected to be not significant as the factor \textit{scenario} was not found to be statistically significant for adjusting aural quality in the MANOVA tests of the previous section. All the other parameters were found statistically significant (reject the $H_0$).

The least square regression estimates of the $\mathbf{M_{2}}$ model are given in Table~\ref{tbl:M2_reg_coefficients}, and the coefficients of determination $R^{2}$ and $R^{2}_{\text{adj}}$ for the goodness of fit for model $\mathbf{M_{2}}$ are given in Table~\ref{tbl:M1_deter_coefficients}.

\begin{table}[!h]
\centering
\renewcommand*{\arraystretch}{1.5}
\begin{tabular}{l rrrr}\hline
\cline{1-5}
Sense & $\hat{\beta_{i}}$ & $\hat{\beta_{b}}$ & $\hat{\gamma_{\text{C}}}$ (Car) & $\hat{\gamma_{\text{K}}}$ (Kitchen)\\
\hline
Smell & $-1.35$ & $0.04$ & $0.72$ & $-1.10$ \\
Visual & $87.33$ & $-0.25$ & $-7.29$ & $\ast$ \\
Audio & $4.72$ & $0.32$ & $\ast$ & $\ast$ \\ \hline
\end{tabular}
\caption{Least squares regression estimates of the multivariate model $\mathbf{M_{2}}$. The asterisks show the parameters that are not statistically significant (not reject the $H_{0}$ hypothesis) and their estimators were excluded from the statistical model.}
\label{tbl:M2_reg_coefficients}
\end{table}

\section{Experiment III: Model Validation}\label{sec:Validation}
The performance of the model was validated in an experimental study which included the use of untested budget sizes and a new untested scenario to test the performance of the model with inputs other than the ones used for the construction.

\subsection{Design}
The validation study follows the same guidelines as the main experimental study. Three scenarios were used, namely, Car, Kitchen and Kitti. Kitti was the new scenario for the validation stage (this was only used for training previously). Three budgets were used in this experiment, each one different from the previously used budgets. These were the midpoints between $B_{1}$ and $B_{2}$, $B_{2}$ and $B_{3}$ and $B_{3}$ and $B_{4}$. These are denoted as $NB_{1}$, $NB_{2}$ and $NB_{3}$. All the materials of the main study, were also used for the validation stage with no change. The Kitchen scenario was used for training. A total of six people participated and none of these participants were familiar with the objectives of the experiment as they had not taken part in the previous experiments. We test both $\mathbf{M_{1}}$ and $\mathbf{M_{2}}$, and for the scenario specific coefficients for $\mathbf{M_{2}}$ we used the values from the Car scenario based on the rationale that the smell source has a similar configuration to the Car scenario.

\subsection{Results}
$\mathbf{M_{1}}$ and $\mathbf{M_{2}}$ were compared with the data collected from the validation study to investigate their performance against real human preferences.

\begin{figure*}[!h]
  \centering
  \includegraphics[scale=0.165]{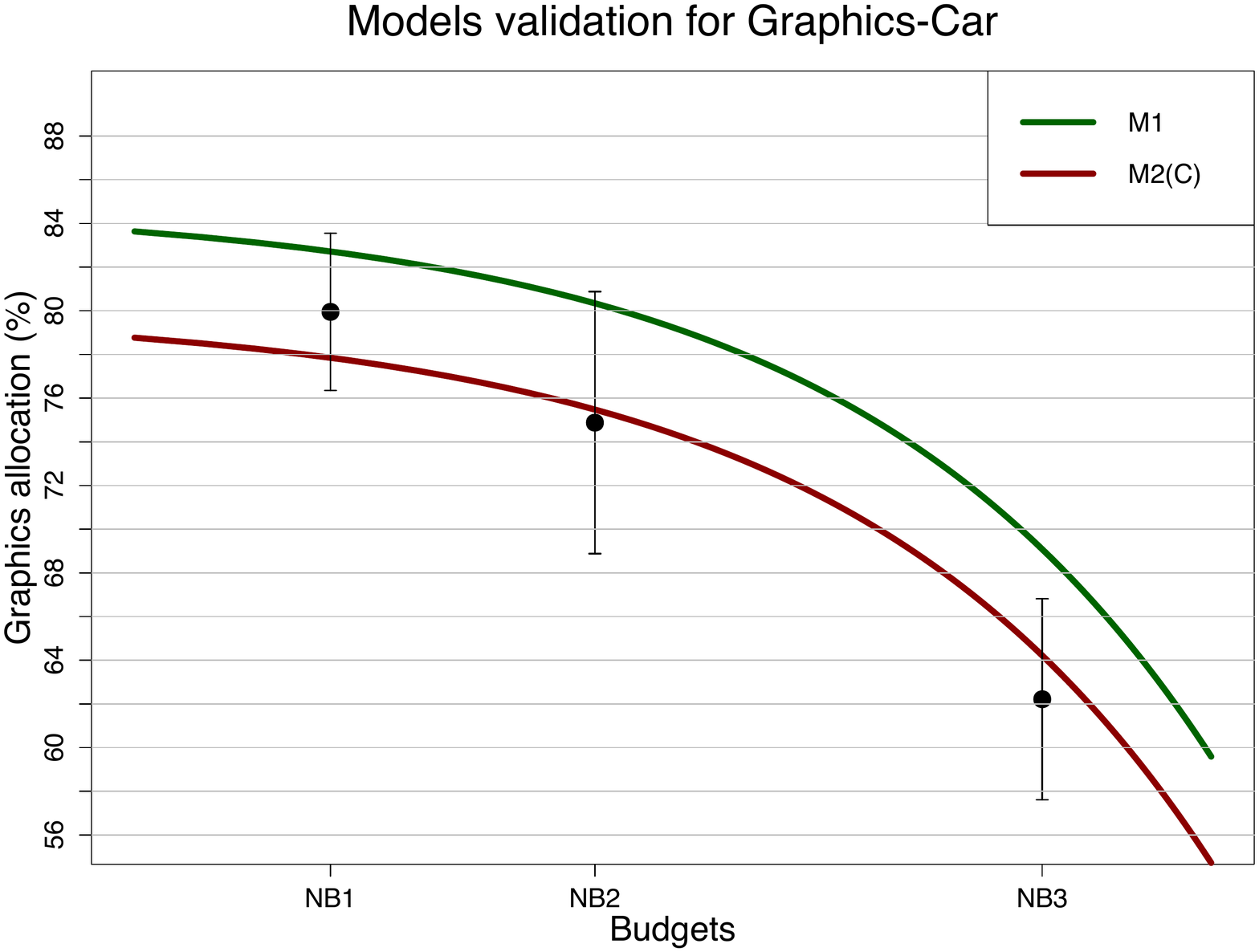}
  \hspace{8.00mm}
  \includegraphics[scale=0.165]{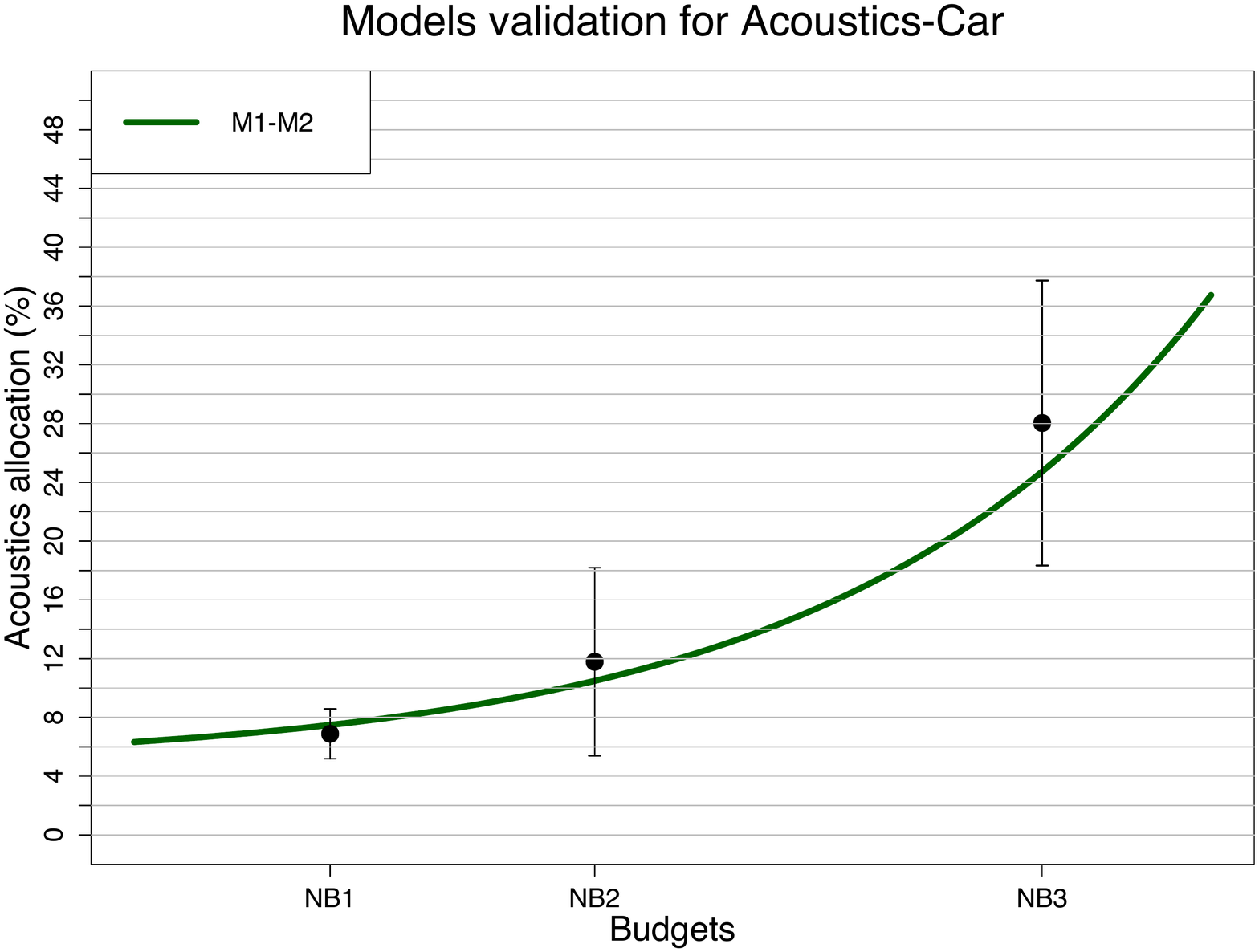}
  \hspace{8.00mm}
  \includegraphics[scale=0.165]{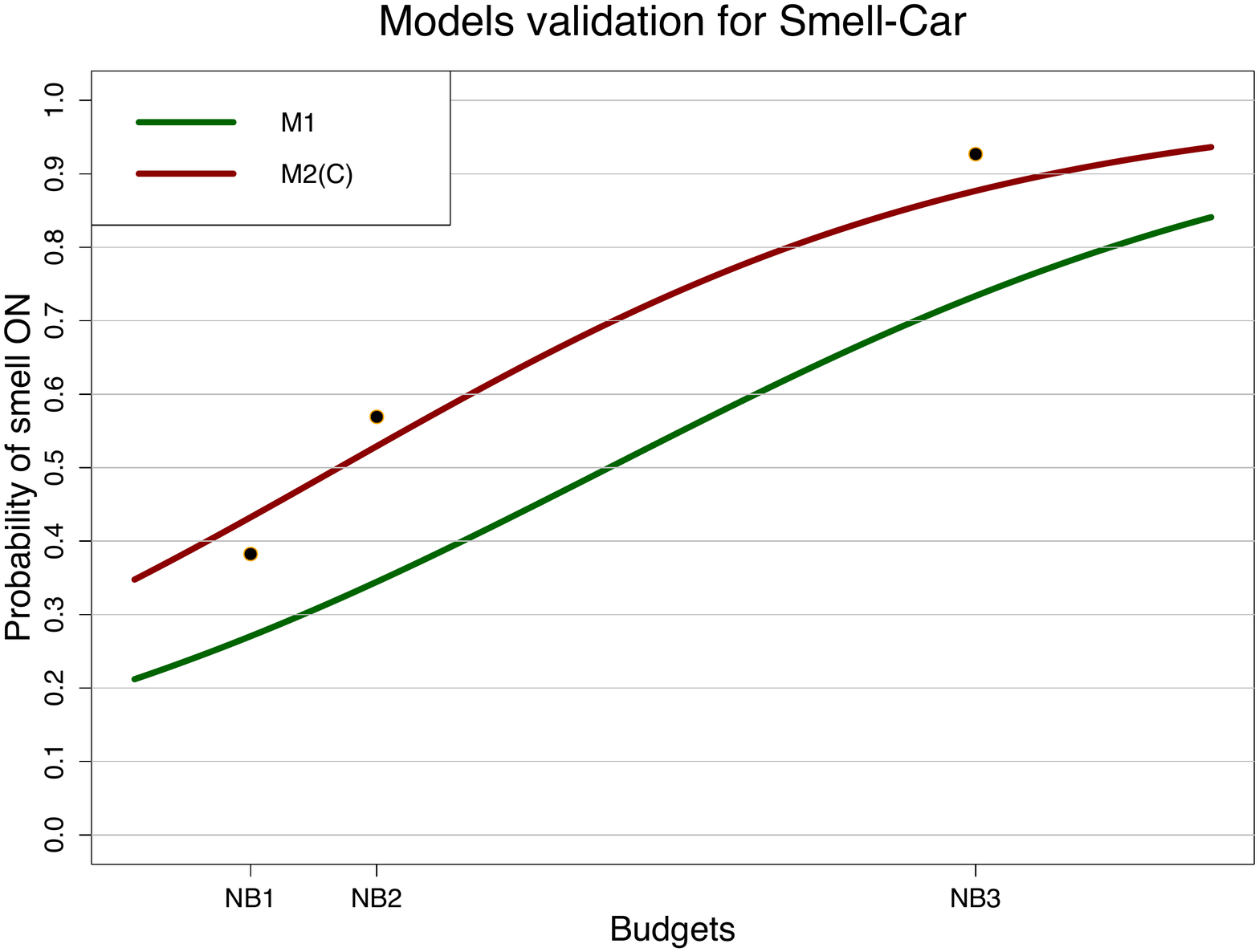}

  \includegraphics[scale=0.165]{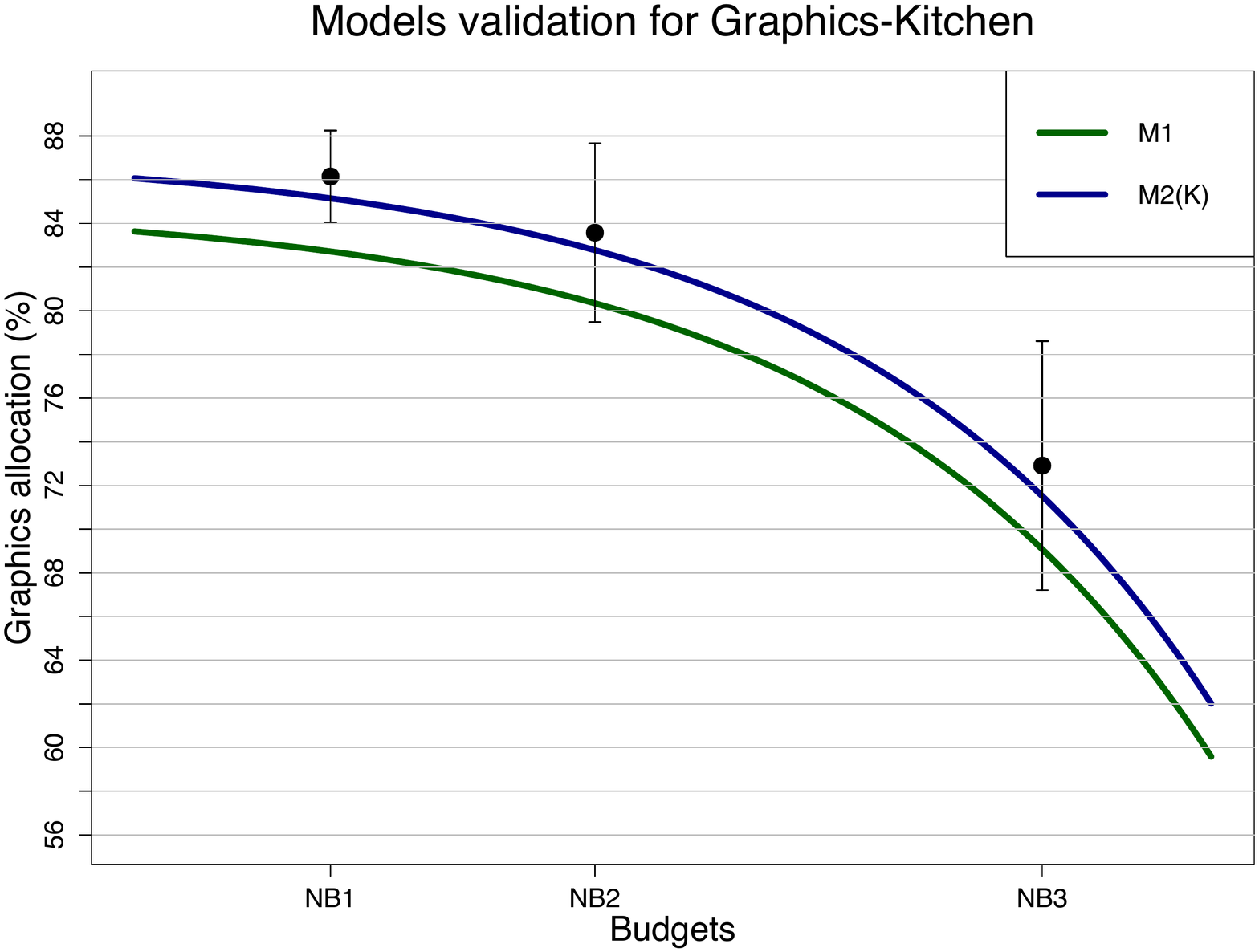}
  \hspace{8.00mm}
  \includegraphics[scale=0.165]{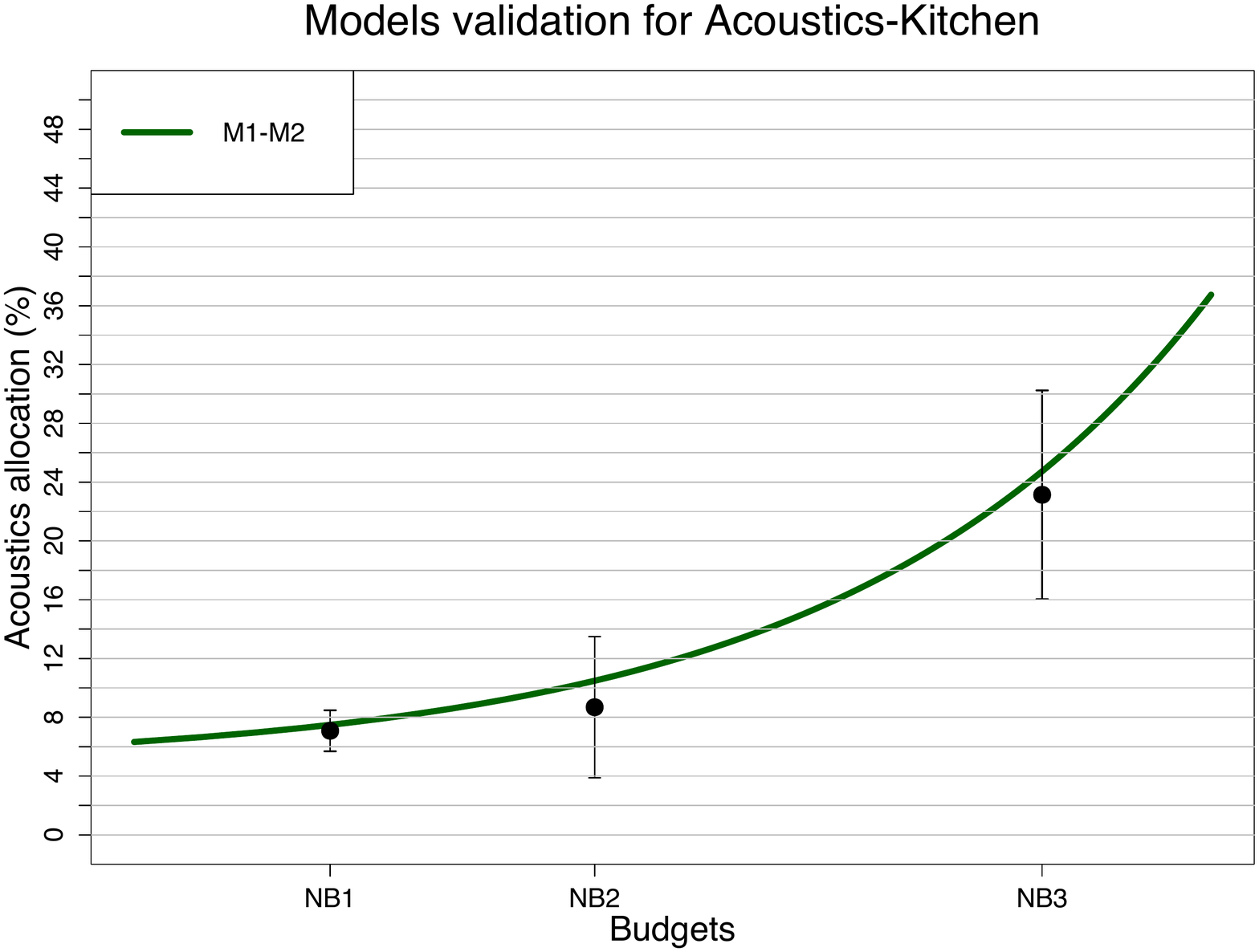}
  \hspace{8.00mm}
  \includegraphics[scale=0.165]{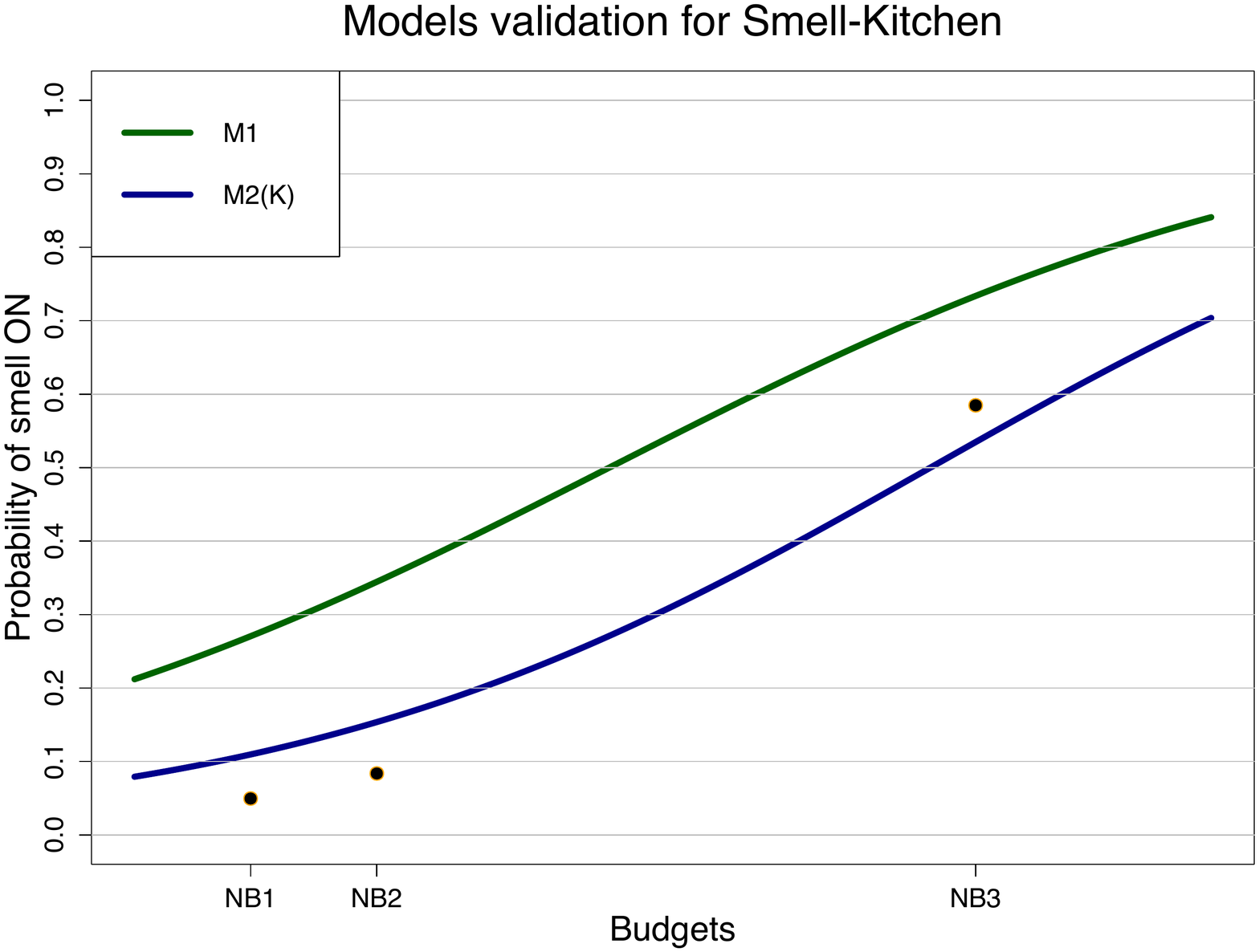}

  \includegraphics[scale=0.165]{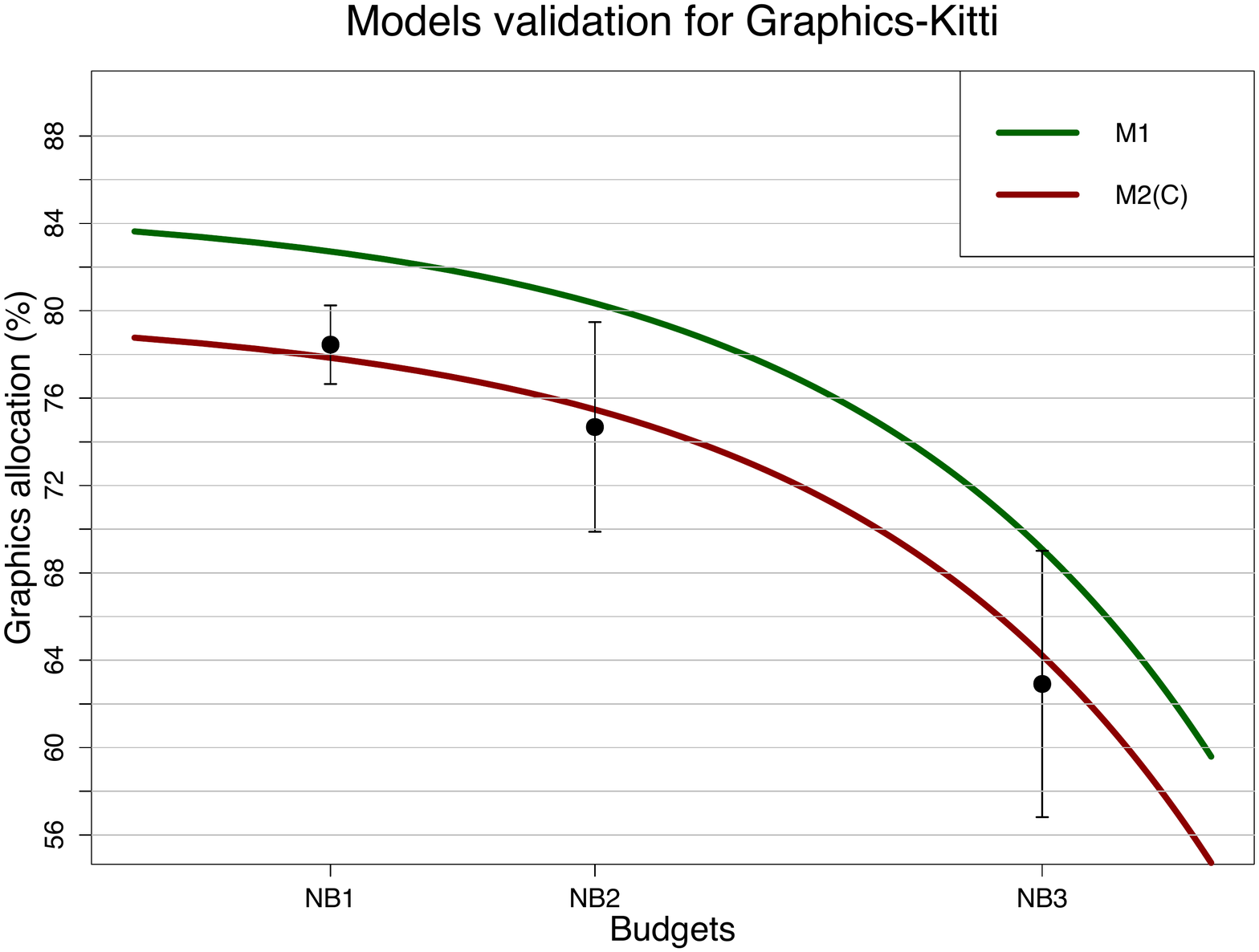}
  \hspace{8.00mm}
  \includegraphics[scale=0.165]{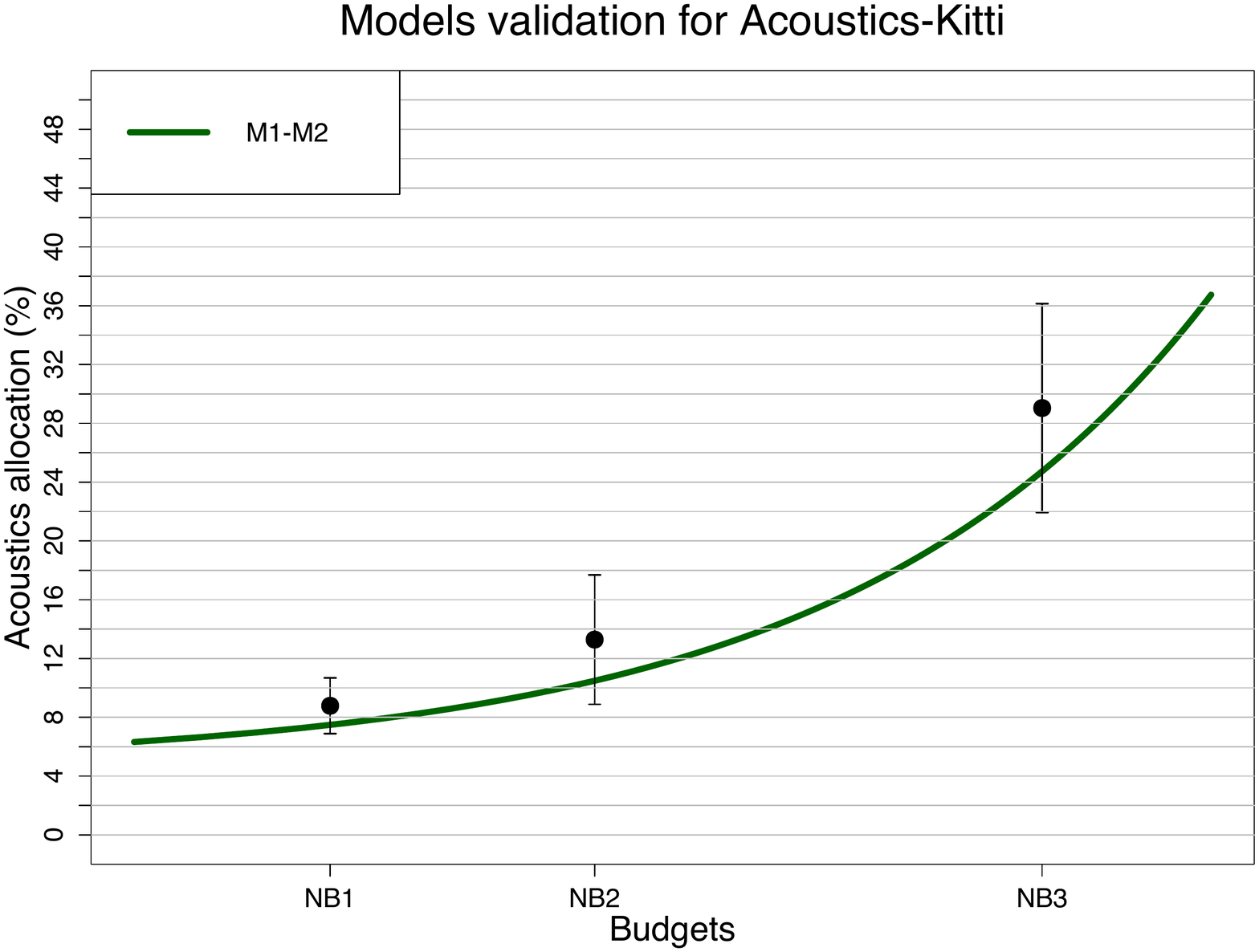}
  \hspace{8.00mm}
  \includegraphics[scale=0.165]{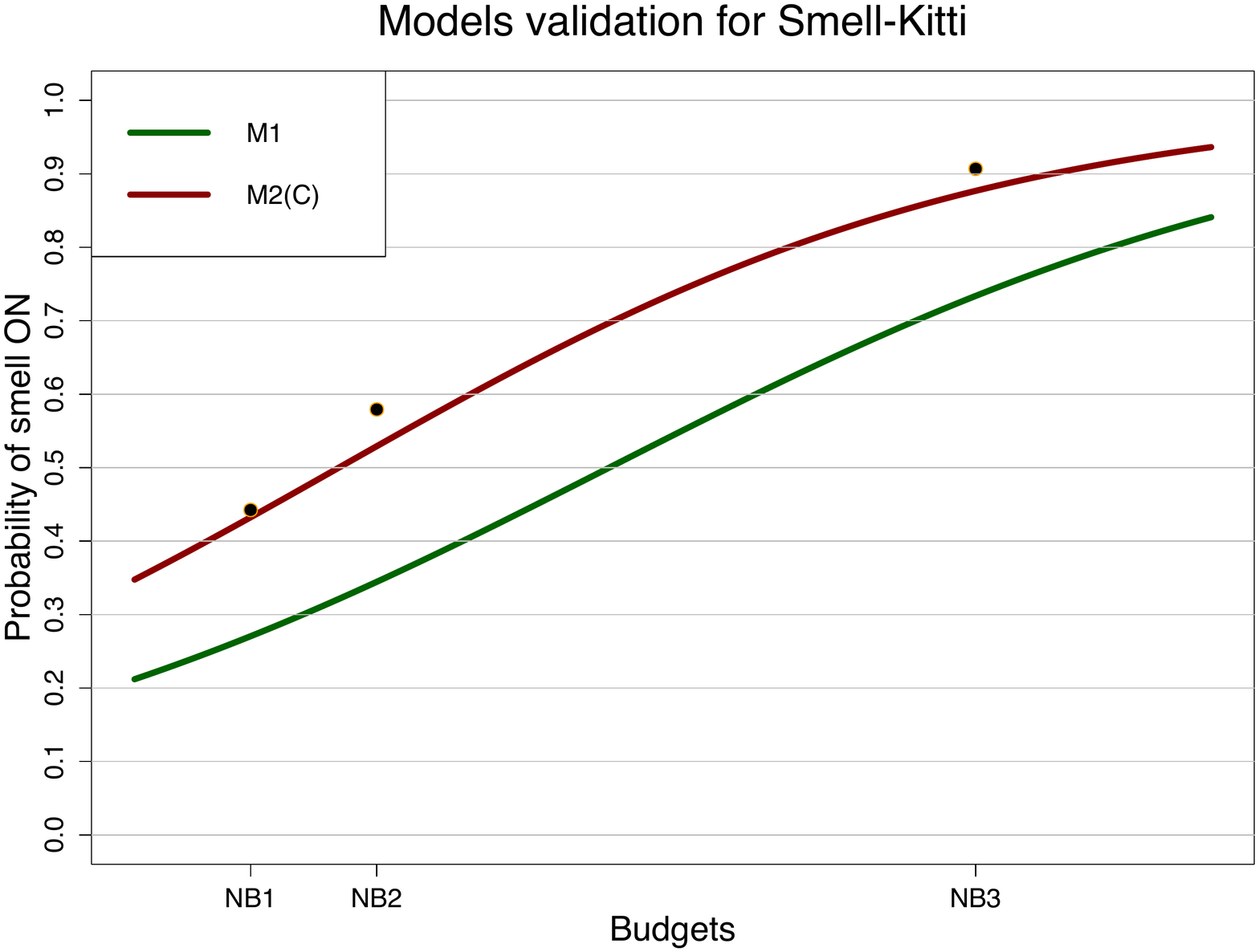}

  \caption{Left and Middle: Means and confidence intervals for the three scenarios for graphics and acoustics respectively. The green curve shows $\mathbf{M_{1}}$ and the red shows $\mathbf{M_{2}}$. The (K) or (C) notations of $\mathbf{M_{2}}$ denote coefficients from the Kitchen or Car scenario as used in $\mathbf{M_{2}}$. Log-spacing was applied to the $x$-axis for better visualisation. Right: Proportion of times smell was preferred on for each of the test budgets across the three scenarios. The green and red curves are the probability estimations to turn the smell on for $\mathbf{M_{1}}$  and $\mathbf{M_{2}}$ respectively.}
  \label{fig:car_validationE3vas}
\end{figure*}

Figure~\ref{fig:car_validationE3vas} shows the means and confidence intervals for results for all scenarios and modalities.
The model estimations are also depicted as colored curves for comparison. For graphics, the average error for absolute difference for the $\mathbf{M_{1}}$ is $5.8\%$ while for $\mathbf{M_{2}}$ is $2.3\%$. For acoustics, $\mathbf{M_{1}}$ and $\mathbf{M_{2}}$ give an average error of absolute difference of $3.4\%$. For small budgets ($NB_{1}$) both $\mathbf{M_{1}}$ and $\mathbf{M_{2}}$ predict equally well for graphics quality. As the budget size increases, model $\mathbf{M_{2}}$ gives more accurate estimations.

For the Kitchen scene, for graphics, the average error for absolute difference is $4.4\%$ and $1.6\%$ for $\mathbf{M_{1}}$ and $\mathbf{M_{2}}$ respectively. For acoustics, $\mathbf{M_{1}}$ and $\mathbf{M_{2}}$ give an average error of absolute difference of $2.3\%$. For smell, at small budget sizes people prefer a low quality audio while smell is not set on. For the test budgets $NB_{1}$ and $NB_{2}$, $\mathbf{M_{1}}$ and $\mathbf{M_{2}}$ overestimate for audio and smell while they underestimate for graphics.


For the new Kitti scenario, whose data was \emph{not} used to build $\mathbf{M_{1}}$ or $\mathbf{M_{2}}$, for graphics the average error for absolute difference is $1.6\%$ and $5.8\%$ for $\mathbf{M_{1}}$ and $\mathbf{M_{2}}$ respectively while for acoustics $\mathbf{M_{1}}$ and $\mathbf{M_{2}}$ give an average error of absolute difference of $4.9\%$ indicating that participants' need for acoustic quality was significantly increased compared to the estimates given by either $\mathbf{M_{1}}$ or $\mathbf{M_{2}}$.




\section{Discussion}\label{sec:discussion}
The results yield a number of potentially useful findings related to the way humans tend to allocate resources in tri-modal VEs. The co-existence of more than two senses in the experimental set-up does not affect humans' general trend to devote the majority of their budget for visual quality improvements. For the scenarios used and the smaller budget sizes, people's priority is to obtain a clear visual stimulus while compromises are made to both the audio quality and the addition of smell. As more resources become available, participants prefer an approximately balanced distribution of resources while the frequency with which smell was added was significantly increased.

As far as the effect of the scenario is concerned, it is clear that the four scenarios affect people's decision to enable the release of smell impulses. The scenario selection was also important for the visual allocation preferences while it had no effect on the percentage devoted to aural quality improvements. Participants did not show significant differences in acoustic resource allocation across the scenarios.

The validation stage shows evidence that the statistical model provides allocation estimates of different accuracy, especially in the case of the visual percentages. $\mathbf{M_{1}}$, the generic version of the model, gives relatively accurate predictions but it was outperformed by model $\mathbf{M_{2}}$ in almost all the combinations of budget-scenario test conditions. The results of the validation study indicate also that both $\mathbf{M_{1}}$ and $\mathbf{M_{2}}$ give relatively accurate estimations for the three different validation budgets and scenario.
When using $\mathbf{M_{2}}$, results are improved at the cost of scene dependent parameters, which currently require experiments to determine. This is an area of future work, as it is likely that these parameters can be predicted as a function of the multi-modal content of the scene.

The delivery of the visual stimuli as static images is a limitation of this experimental study as it restricts the number of possible applications of the proposed estimation model. Both the auditory and olfactory stimuli have an inherent temporal dimension (duration of the audio track/smell burst) that is not exploited with the use of static visual images. However, the experimental methodology can provide useful insights on extending to VEs with animations or interactive environments.

Another possible limitation is the use of the binary (on/off) smell stimuli. It can be said that the two-level methodology for delivering smell impulses restricts user's available options but it is not clear whether (and how) a hypothetical multiple quality scale for smell would benefit user's virtual experience, particularly with current olfactory hardware limitations, as shown in Experiment I. The existence of multiple smells, on the other hand, is considered more important in a multi-sensory VE and can admittedly increase the level of immersion as it is closer to what happens in the real word. To that end, the JND estimation methodology can be applied in arbitrary number of VOCs or combinations of them.


\section{Conclusion}\label{sec:conclusion}
This paper presented a series of experimental studies with the ultimate goal of tri-modal resource allocation for VEs.
Unlike previous work, which focused predominantly on audio-visual cross modalities, participants were given the opportunity to release an olfactory stimulus. Based on a first experiment, the sense of olfaction included two different quality levels (smell on/off) while smell impulses were delivered to the user using an olfactory display.
This was used to develop a model for resource allocation for tri-modal virtual environments, which has been validated in a third experimental study.
Future work will consider dynamic tri-modal VEs, where the temporal nature of visual, auditory and olfactory stimuli will be investigated. Additionally, we intend to explore whether the allocation of computational resources is impacted by user movement in the VE.

%


\bibliographystyle{abbrv-doi}
\bibliography{references}
\end{document}